\documentclass[sigconf]{acmart}
\usepackage{framed}
\usepackage{soul}

\renewenvironment{framed}[1][\hsize]
{\MakeFramed{\hsize#1\advance\hsize-\width \FrameRestore}}%
{\endMakeFramed}

\copyrightyear{2026}
\acmYear{2026}
\setcopyright{cc}
\setcctype{by}
\acmConference[DIS '26]{Designing Interactive Systems Conference}{June 13--17, 2026}{Singapore, Singapore}
\acmBooktitle{Designing Interactive Systems Conference (DIS '26), June 13--17, 2026, Singapore, Singapore}
\acmDOI{10.1145/3800645.3813031}
\acmISBN{979-8-4007-2563-0/2026/06}

\begin{document}

\title{Putting a Face to the Issue: Fostering User Empathy of Open Source Software Developers With PersonaFlow}

\author{Boniface Bahati Tadjuidje}
\email{bahati-tadjuidje.boniface@polymtl.ca}
\orcid{0009-0004-8386-9664}
\affiliation{%
  \department{Department of Computer and Software Engineering}
  \institution{Polytechnique Montreal}
  \city{Montreal}
  \state{Quebec}
  \country{Canada}
}

\author{Jin L.C. Guo}
\email{jin.guo@mcgill.ca}
\orcid{0000-0003-1782-1545}
\affiliation{%
  \department{School of Computer Science}
  \institution{McGill University}
  \city{Montreal}
  \state{QC}
  \country{Canada}
}
\author{Jinghui Cheng}
\email{jinghui.cheng@polymtl.ca}
\orcid{0000-0002-8474-5290}
\affiliation{%
  \department{Department of Computer and Software Engineering}
  \institution{Polytechnique Montreal}
  \city{Montreal}
  \state{QC}
  \country{Canada}
}

\begin{abstract}
Open-source software (OSS) developers often struggle to understand and respond to user context, while existing tools, such as issue trackers (for handling bugs, requests, and feedback), largely focus on technical discussion. Although personas could help, limited resources and UX expertise make them hard to scale. We present PersonaFlow, a tool that generates editable user personas from OSS repository artifacts and integrates them alongside issue reports. In a user study with 13 OSS developers, most reported shifts in how they understood users, and more than half modified their responses by adding empathetic language, tailoring explanations, or raising priority ratings. We found two pathways to this change: some connected emotionally to personas as people, while others used them pragmatically for triaging. Both appeared to lead to more user-centered behavior. We contribute design implications for persona-based tools relevant to OSS and other contexts where efficiency-driven systems or workflows obscure valuable human elements.
\end{abstract}



\maketitle

\section{Introduction}
Open-source software (OSS) development relies on social coding platforms to build and sustain its communities. Tools and platforms like issue trackers~\cite{Arya2019} and discussion forums~\cite{hellman2022characterizing} offer the promise to allow developers and users to have direct communication to address user needs. However, in reality, these channels capture only a self-selecting subset of users (typically those with enough technical vocabulary to articulate problems) and rarely reveal underlying goals, frustrations, or the human context behind requests~\cite{argulens}. To aggravate this problem, OSS communities often lack the necessary resources and expertise to conduct user research or follow typical UX practices~\cite{wang2022how}. As a result, OSS developers and maintainers often struggle to understand their diverse user base, defaulting to technical-first prioritization that treats issues (bug reports, feature requests, and other project-related discussions) as puzzles to solve rather than people to help~\cite{feller_perspectives_2005,wang2022how}.

This resource gap creates a persistent empathy challenge in OSS communities. Research has documented widespread patterns of dismissive responses in communication channels, unwelcoming behavior toward newcomers~\cite{steinmacher2019let}, and communication failures that escalate into toxic and uncivil exchanges~\cite{miller2022did, ferreira2021shut}. These are not isolated incidents but symptoms of a deeper problem: the information environment of OSS development provides rich technical data (labels, milestones, code references) but almost no user context (who is affected, how severely, what their goals are). This is not a matter of individual disposition but of information design. Developers simply lack the awareness triggers needed to empathize with users they never meet~\cite{gunatilake2024enablers}. Personas---fictional representations of a product's target user types---have long been recognized as effective tools for building empathy and grounding design decisions in user needs~\cite{cooper1999inmates, pruitt2003personas}. However, traditional persona creation requires specialized UX expertise and extensive work that most OSS projects simply cannot afford.

To explore how AI-generated personas could support OSS developers in understanding their users, we designed PersonaFlow, a tool that addresses this gap by supporting the generation of user personas from repository artifacts and integrating them directly into the issue management workflow. From a repository URL, PersonaFlow can generate diverse personas grounded in the actual project context through analysis of README files, documentation, and issue reports. Users can also provide any additional documents or resources to complement the project repository. Moreover, PersonaFlow maps these personas to individual issues in the issue tracker with confidence scores and explanations, showing developers not just what is broken about the software but \textit{who is affected by this issue and how}. Developers can review, edit, merge, or regenerate personas at any point in the workflow, maintaining necessary oversight. Finally, an analytics dashboard is provided to help teams identify when certain user types face disproportionate issues. By embedding personas alongside the issue tracker that developers regularly use to manage and prioritize their work, PersonaFlow provides the context needed to prompt OSS developers to reflect on the impact on users.

In a user study with 13 OSS developers, participants used PersonaFlow on repositories they are familiar with to explore AI-generated personas, respond to issues before and after persona exposure, and provide feedback on the tool's features. Rather than testing specific hypotheses, our goal is to explore how developers engage with AI-generated personas in the context of issue handling and identify design considerations in this space. Our findings revealed that most of the participants reported shifts in how they understood users, and more than half modified their issue responses after persona exposure---they added empathetic language or tailored explanations in their replies and raised priority ratings based on user impact. We identified two distinct pathways through which personas nudged our participants to react to user needs. Some developers connected emotionally to personas as real people (empathy-driven), while others used them pragmatically for efficient issue resolution (utility-driven). These findings suggest that persona tools can benefit diverse developer orientations. Based on these findings, we contribute design implications for AI-assisted persona tools that can help OSS developers better understand and support their users. These findings can also extend to other work contexts where efficiency-oriented systems or workflows abstract away the human elements important for perspective shifting.

\textbf{Researcher Positionality.}
Our understanding of the OSS development workflow, as well as our approach in this study, is informed by our prior experience researching related topics in the context of HCI and software engineering, as well as by our direct participation in OSS projects. In particular, the first author is an active OSS contributor, currently maintaining multiple repositories. This provides the research team firsthand experience with issue handling and user interactions, and how this experience interplays with developers' mindsets. The remaining authors had extensive prior research experience studying the aspects of usability, design, inclusiveness, and sustainability in OSS and collaborative work. All authors have had prior training and varied experience in interaction design and UX research. Overall, this positionality shaped how we approached the system design and how we collected, analyzed, and interpreted the user study data. 

\section{Related Work}
Our work builds upon prior research on the empathy challenges during software development, the use of personas as an empathy-building mechanism, and the recent emergence of generative AI for persona creation.

\subsection{The Empathy Gap in Distributed Software Development}
The rise of distributed software development, accelerated by globalization and remote work technologies, has fundamentally transformed how software teams collaborate across geographical, temporal, and cultural boundaries~\cite{carmel2001work}. Open-source software (OSS) development is a prominent example of distributed software work, yet it presents unique challenges in managing contributions across diverse contexts. Through a series of studies,  \citet{steinmacher2019let} identified 57 distinct barriers that newcomers face in OSS, with communication issues, including impolite responses and unwelcoming behavior, representing a significant category. While this communication breakdown happened between developers, it reflects deeper challenges in understanding and responding to others' needs in remote and distributed development environments. Building on this, \citet{gunatilake2024enablers} found that while developers require specific ``awareness triggers'' to activate empathy toward users, OSS environments systematically lack these triggers due to limited direct user contact and asynchronous communication patterns, creating impediments to empathy building.

Recent work has shed more light on the barriers between developers and users and their consequences~\cite{Hellman2025}. Developers and users often speak different ``languages'' even when discussing the same issues~\cite{hellman2022characterizing}, compounding the empathy challenge. Such communication failures can escalate into confrontational conflicts, as revealed by widespread and nuanced incivility and toxicity across OSS platforms. For example, research has detailed patterns of incivility, such as frustration and name-calling, within Linux Kernel code review discussions \cite{ferreira2021shut}, while analysis of toxic discussions on GitHub reveals them to be empathy breakdowns at scale \cite{miller2022did}. However, understanding and managing these interactions is complex; a study of GitHub's ``locked issues'' feature found that many discussions labeled as ``too heated'' by maintainers did not actually contain uncivil content, pointing to inconsistencies in how communities identify and handle conflict \cite{ferreira2022how}. More critically, \citet{sin2021digital} introduced the concept of ``Digital Design Marginalization'', demonstrating how non-inclusive designs push certain users away from digital services with lasting social consequences. This marginalization is particularly evident in OSS: \citet{balali2018newcomers} showed how project structures encode systematic biases against women, while \citet{wang2022how} found that OSS contributors struggle with usability concerns due to a predominance of technical thinking that overshadows user perspectives. 

These findings show that the empathy gap in distributed software development is not an isolated issue, but a systemic problem rooted in its information environment and priorities manifested in existing tools. Our work addresses this challenge by proposing a tool-based intervention that surfaces user context at the point of developer-issue interaction, making it easier for developers to consider who is affected by the issues they handle.

\subsection{Personas as Empathy-Building Tools}
Personas, a user-centric design tool popularized by Alan Cooper~\cite{cooper1999inmates}, are often used to bridge different communities of practice, such as developers, designers, and project managers~\cite{marsden2016stereotypes}. Aiming to cultivate empathy and engagement with users, personas can be created by designers through user research or co-created with users~\cite{neate2019cocreated}. From their experience at Microsoft, \citet{pruitt2003personas} observed that personas helped teams create ``a strong focus on users'' and provided a ``shared basis for communication.'' Similarly, \citet{matthews2012designers} found that practitioners value personas primarily as practical tools for communication and decision-making, which makes abstract users concrete and discussable.

Despite their effectiveness, the adoption of personas is often hindered by practical constraints. \citet{pruitt2003personas} noted that traditional persona creation, which relies on deep user research, can require months of intensive data collection and analysis, making it prohibitively expensive for many organizations. This challenge is especially acute in resource-constrained OSS projects. These resource limitations can force trade-offs that lead to personas becoming outdated or unused, a phenomenon known as the ``file drawer effect''~\cite{salminen2018personas}. A recent survey of 203 practitioners by \citet{wang2024personas} revealed that while developers recognized the value of personas, practical barriers, including time, cost, and lack of expertise, prevented their regular use in software projects.

Personas also carry risks when poorly implemented. \citet{marsden2016stereotypes} warned that personas can reinforce stereotypes and biases if not grounded in authentic user data, potentially causing more harm to users than good. \citet{bennett2019promise} later offered a deeper critique, cautioning that personas can inadvertently ``sustain the very difference that it may seek to overcome'' if they present users as fundamentally ``other'' rather than fostering genuine understanding. The high cost and potential pitfalls of traditional methods highlight the need for more accessible and reliable approaches that can scale to the needs of distributed software development.

\subsection{Persona Creation with Data-Driven Automation}
Recent advances in generative AI have opened new possibilities for creating personas at scale, potentially addressing the resource constraints that limit traditional approaches. The field has seen explosive growth, with recent HCI research demonstrating both the promise and challenges of this approach. Systems like PersonaCraft~\cite{jung2025personacraft} have demonstrated that AI can generate credible and complete persona profiles from user data, achieving quality ratings comparable to traditionally-created personas. \citet{zhang2024auto} showed that AI-generated personas can effectively support design education, helping students develop user-centered thinking.

The ability to assess the quality of AI-generated personas is a critical aspect of their development. \citet{salminen2024deus} investigated 450 LLM-generated personas and found that they are rated differently by subject matters and UX researchers. \citet{salminen2020persona} developed and validated the Persona Perception Scale, a framework for evaluating personas on dimensions of credibility, empathy, and likability. Recent research into human-AI collaboration has focused on \textbf{how} to create high-quality personas. \citet{shin2024humanai} found that combining the strengths of both humans and AI produces personas that are more representative and empathy-evoking than those created by either humans or AI alone. In this collaborative model, the AI handles pattern recognition and synthesis across large datasets, while humans provide contextual understanding and quality control.

While modern LLMs may address many technical shortcomings, a fundamental challenge remains: AI-generated personas, despite their increasing sophistication, still struggle to capture the authentic complexity of human users. When personas lack grounding in real user expressions and contexts, they feel hollow even when they meet technical quality measures~\cite{probster2018perceptions}. Recent work on human-AI iteration also reveals a broader challenge on trust: \citet{vereschak2024trust} found that trust in AI systems is strongly influenced by human stakeholders rather than system features alone, while \citet{ma2023trust} demonstrated that humans struggle to appropriately calibrate their trust in AI during decision making. 

PersonaFlow builds upon this line of work by considering both challenges. It grounds persona generation in human-authored project artifacts, including documentations and issue discussions, to preserve user-targeted voices, while supporting control and trust calibration through confidence indicators, explicit explanations, and extensive editing capabilities. Moreover, with the recognition that automatically generated personas will always risk misrepresenting users, our work focuses on how persona-based tools can incorporate automation in a human-centric workflow for scale, while providing OSS developers the much-needed ``awareness triggers''~\cite{gunatilake2024enablers} for kick-starting their user-centric behavior.

\section{PersonaFlow's Design and Implementation}
The design and implementation of PersonaFlow followed an iterative, user-centered approach. In the following sections, we describe our design goals (DGs), the design and implementation process, the final user interaction design of PersonaFlow, as well as its implementation details.

\subsection{Design Goals and Considerations}
To ensure that our tool effectively addresses the needs of OSS developers, we identified the following four key goals and considerations, informed by the literature and our own experience working on OSS projects and tools, to guide our design process.

\textbf{DG1: Providing Insights into User Impact.} Effective decision-making requires situation awareness~\cite{endsley1995situation}. Research shows that during their regular workflow, OSS developers systematically lack the ``awareness triggers'' needed to activate empathy~\cite{gunatilake2024enablers}, often defaulting to technical-first prioritization without being conscious of user context. Personas can address this gap by making user impact visible and actionable~\cite{pruitt2003personas}. We thus aim to incorporate personas in the OSS tool space to promptly provide insights into user impact to support OSS developers.

\textbf{DG2: Low-Barrier Automated Persona Generation.} Traditional persona creation requires considerable effort, specialized expertise, and substantial budgets~\cite{pruitt2003personas}. Practical barriers, such as time, cost, and expertise, may prevent regular persona use and related practices~\cite{wang2024personas}. OSS communities especially lack these resources~\cite{wang2022how}. Recent work shows AI can generate useful personas from existing data~\cite{salminen2024deus}. We thus explore how automated techniques can be leveraged to lower these barriers. 

\textbf{DG3: Workflow Integration and Flexibility.} Prior research highlighted key factors impacting adoption, especially on how compatible the intervention is with existing values, experiences, and practices of practitioners~\cite{palani2022adoption,Shokrizadeh2025}. In software development specifically, automated tools that disrupt established workflows face resistance despite their intended benefits~\cite{wessel2021disturb}, while tools well-integrated into the existing work see higher adoption~\cite{razzaq2024devx}. In our tool design, we emphasize its integration into the existing social computing landscape for OSS communities, including both development tools (e.g., GitHub) and processes (e.g., issue triaging and responding).

\textbf{DG4: Transparency and User Control.} Research on human-AI collaboration reveals that users often struggle to calibrate trust with AI appropriately, leading to over-reliance or under-reliance~\cite{zhang2020effect, duan2025trusting}. Transparency and explanations that reveal AI limitations and uncertainty, rather than those designed to persuade acceptance, can foster appropriate trust~\cite{mehrotra2023integrity}. At the same time, AI-generated personas risk reinforcing stereotypes through biased representations of demographic groups~\cite{salminen2024deus}. These findings suggest AI persona tools must provide uncertainty indicators (e.g., confidence scores that signal when AI is less certain), expose AI reasoning for critical evaluation rather than blind acceptance, and afford human control to validate and override AI outputs. 

\subsection{Design and Implementation Process}
\label{sec:design_process}
Guided by these design goals, we followed an iterative process to design and implement PersonaFlow. The design process involved multiple stages of ideation, prototyping, feedback collection, and refinement, ultimately resulting in a comprehensive web-based tool for persona-driven OSS development.

\textbf{Initial Ideation.} 
Our initial ideation centered on a key insight from DG3: personas alone, however well-crafted, provide limited value if they sit outside of OSS developers' actual work. Prior research shows that personas can remain unused when they exist as static artifacts separate from daily workflows~\cite{salminen2018personas}, but their value shines when integrated as communication tools for decision-making~\cite{matthews2012designers}. In OSS contexts, issue handling represents the central activity for decision making and the primary interface between developers and users---issues are where user problems surface, prioritization decisions are made, and the consequences of those decisions become visible~\cite{steinmacher2019let}. We thus decided to design PersonaFlow around this workflow, making personas actionable by connecting them directly to the issues where developers already engage with user needs.

When initially ideating PersonaFlow, we explored two options for its overall architecture, also considering DG3's emphasis on workflow integration: (1) a browser extension that would overlay persona information directly onto the GitHub web interface, versus (2) a standalone tool that integrates with GitHub through its API. During our examination of those two options, we realized a tension in our tool design: persona-driven development involves both reactive work (e.g., triaging issues, quick prioritization) that can benefit from lightweight tool augmentation, and reflective work (e.g., analyzing user segments, recognizing cross-issue patterns, refining personas) that can benefit from a dedicated environment. Browser extensions excel at reactive augmentation but risk fragmenting reflective tasks across constrained sidebars and overlays. As a novel tool idea, we saw PersonaFlow's core value in supporting the reflective work. We thus prioritized designing this tool as a coherent, standalone environment, while acknowledging that a browser extension would offer superior workflow integration. Future iterations could explore hybrid architectures combining both approaches.

\textbf{Prototyping and Design Iteration.} Following the architectural pivot, we developed initial design sketches and mockups for the web-based interface, focusing on three key workflows: (1) repository analysis and persona generation, (2) persona management, and (3) issue management with persona associations. Different versions of the sketches and mockups were reviewed iteratively by team members. After several design iterations, we implemented a working version with the core functionalities of the tool.

This process revealed several tensions among our design goals that require careful navigation. The most significant was \textit{automation versus human control}, stemming from the conflict between DG2 (lowering barriers through automation) and DG4 (maintaining user agency). Fully automated persona generation would minimize effort but risk generating invalid or irrelevant content; manual creation would ensure relevance but recreate traditional barriers (e.g., demand on resources and expertise). We resolved this by emphasizing a ``generate-then-refine'' approach where AI produces initial personas that developers can edit, merge, or delete. Similarly, \textit{information richness versus cognitive load} arose from pursuing DG1 (providing rich user insights) while maintaining usability to developers who may not be familiar with the concept of persona. We addressed this through progressive disclosure, showing summary cards with expandable details. Finally, \textit{confidence shortcut versus explanation} directly emerged from DG4's trust calibration concerns when we design features to map personas with issues. Specifically, quantitative confidence metrics offer efficient shortcuts but risk misinterpretation, while explanations of the linkage between issues and personas promote calibrated understanding at the expense of additional cognitive effort. We settled on providing both visual confidence indicators and short explanatory reasoning to navigate this tradeoff.

\textbf{Preliminary User Study and Further Improvements.} To further iterate the design, we conducted a preliminary user study with two OSS developers (not included in the final study), each lasting approximately one hour. Participants used the tool with their own repositories and provided feedback on workflows, interface clarity, and persona usefulness.

These studies surfaced several issues that we addressed accordingly. For example, persona generation was originally handled asynchronously---users will be notified after the generation process is finished. Participants found this process opaque, uncertain whether the system was progressing or stalled. We thus added clearer indicators with stage-by-stage feedback for the generation progress. Both participants also wanted to refine personas right after generation, instead of going into a separate screen to inspect and edit them. So, we improved editing capabilities to better support the ``generate-then-refine'' workflow. Further, participants wanted to have immediate and interpretable information about how certain the AI was about any association between issues and personas it predicts, leading us to add visual confidence indicators that expose uncertainty. Moreover, participants found common tasks like associating additional personas with issues required too many steps, so we streamlined these workflows to reduce friction. One feature we removed was automatic persona suggestions during issue creation, which participants found interruptive rather than helpful. This reinforced our assumption that workflow integration requires careful timing, not just proximity. Finally, participants noted that some initially generated personas felt generic. We refined the AI prompts by requiring persona goals and pain points to reference specific product features, and by instructing the model to focus on behaviors in the personas rather than demographic assumptions.

The preliminary user study also confirmed that our core design tensions were genuine: participants valued automation but immediately sought editing control, appreciated confidence scores but needed explanatory reasoning to act on them, and wanted rich persona profiles but engaged primarily with summary views. These observations validated our resolution strategies while revealing that DG4 (transparency and user control) required more interface prominence than initially planned. The final version of PersonaFlow reflects these insights, prioritizing visible confidence indicators and low-effort editing over an automated but opaque experience.

\begin{figure*}[t]
\centering
\includegraphics[width=0.95\textwidth]{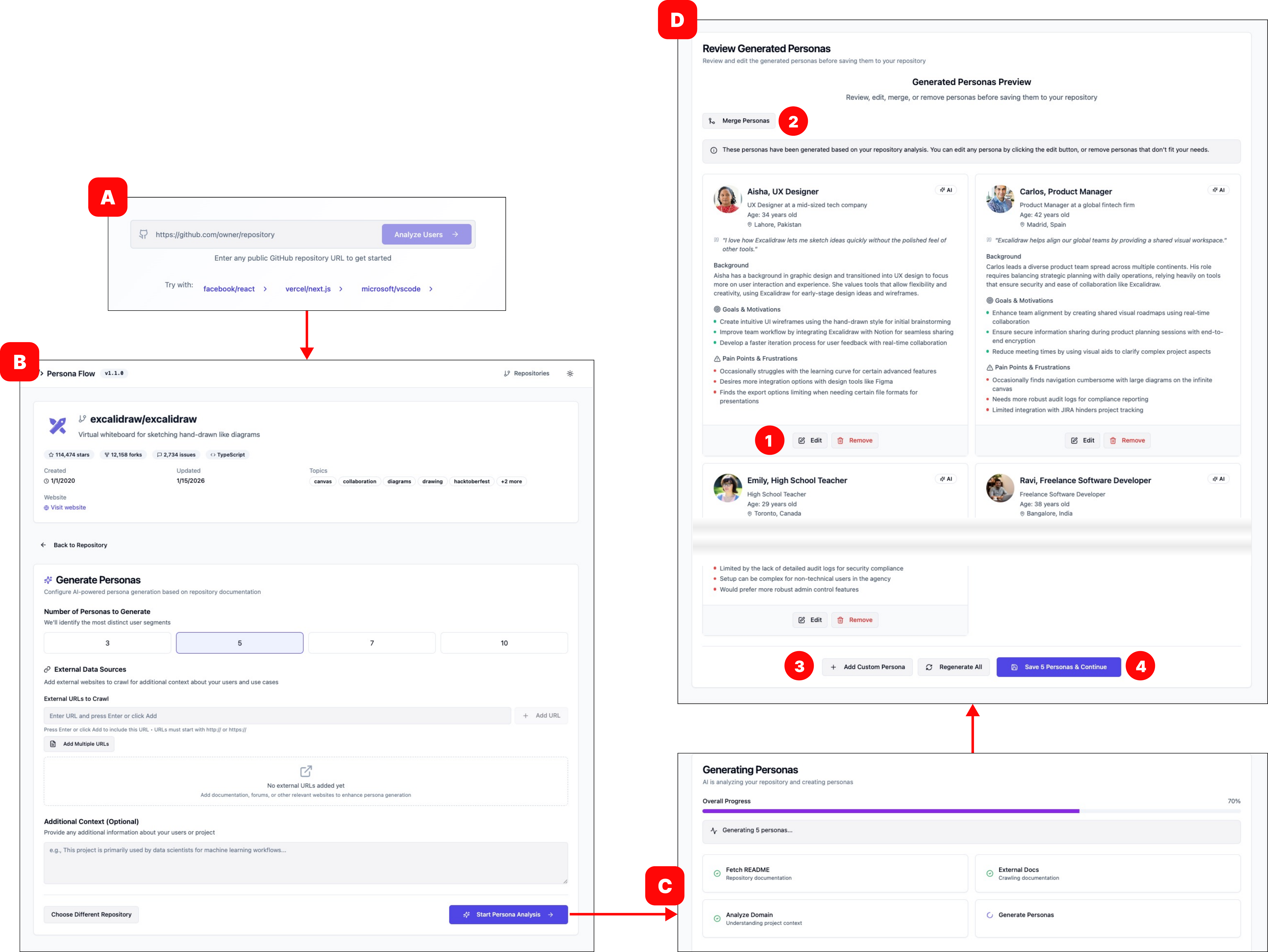}
\caption{Repository analysis and persona generation workflow. Users input the repository URL (A), configure generation parameters (B), track real-time generation progress (C), and preview, edit, and approve the generated personas (D).}
\label{fig:workflow_persona_generation}
\Description{Four connected screenshots showing the persona generation workflow. Panel A shows a text input field for a GitHub repository URL with an ``Analyze Users'' button and suggested repositories. Panel B shows the configuration page for the excalidraw/excalidraw repository, displaying repository metadata (114,476 stars, 12,158 forks, 2,734 issues), a slider to select the number of personas (3, 5, 7, or 10), fields for external URLs to crawl, and an optional additional context text area, with a ``Start Persona Analysis'' button. Panel C shows a progress dialog at 70\% completion with four processing stages: Fetch README, External Docs, Analyze Domain, and Generate Personas. Panel D shows the ``Review Generated Personas'' page with four persona cards (Aisha the UX Designer, Carlos the Product Manager, Emily the High School Teacher, and Ravi the Freelance Software Developer), each displaying a headshot, role, age, location, background, goals and motivations, and pain points. Each card has Edit and Remove buttons (circled 1). A toolbar at the top offers Merge Personas (circled 2), and at the bottom, Add Custom Persona and Regenerate All buttons (circled 3) alongside a ``Save 5 Personas and Continue'' button (circled 4).}
\end{figure*}

\begin{figure*}[t]
\centering
\includegraphics[width=\textwidth]{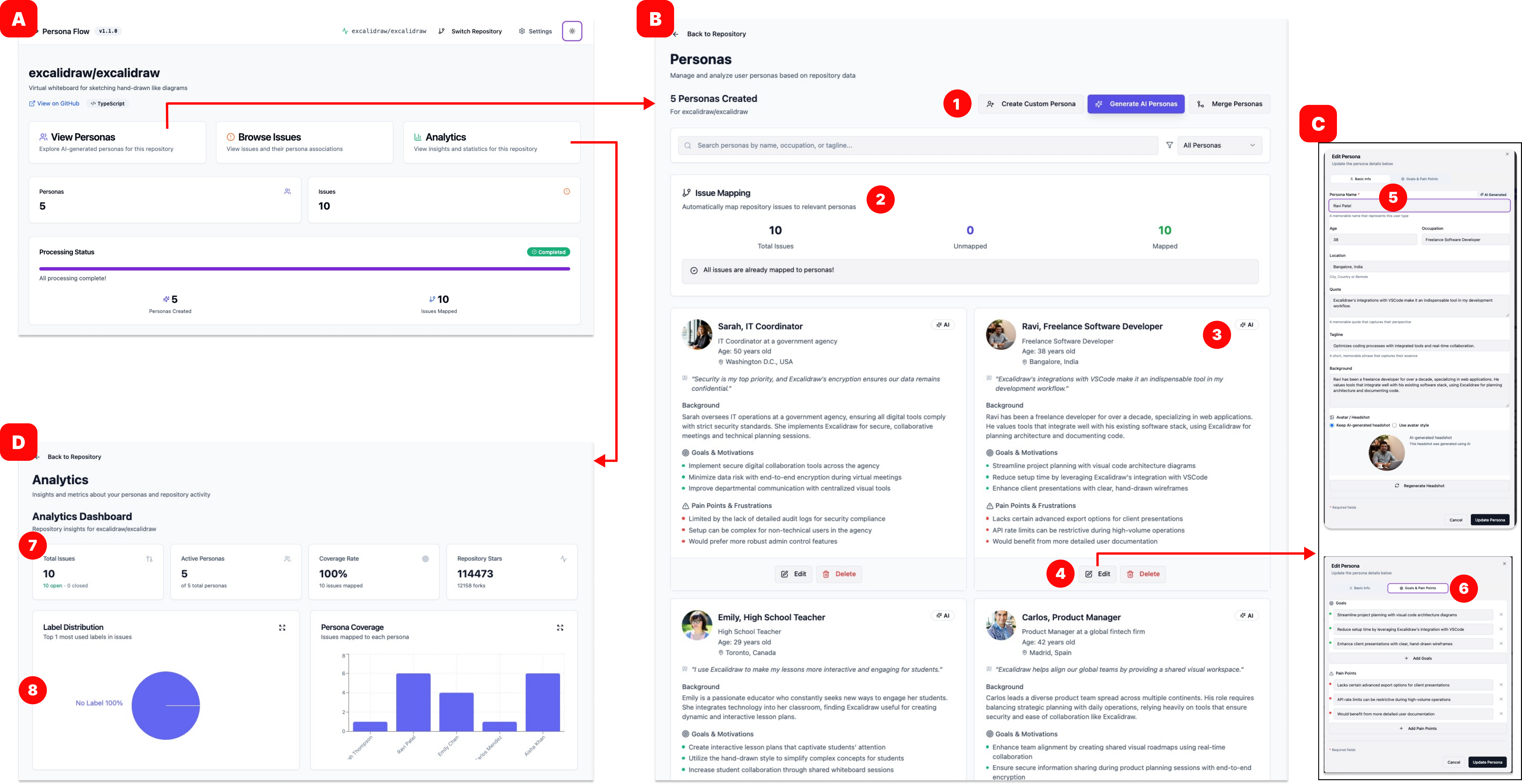}
\caption{Persona management and analytics workflow. The repository dashboard (A) shows three main functions: View Personas, Browse Issues, and Analytics. The View Personas function (B) allows users to inspect current personas, manually create or generate new personas, merge personas, edit personas (C) and inspect issue-persona mapping status. The Analytics dashboard (D) shows summary metrics and visualizations of persona coverage and label distribution.}
\Description{Four connected screenshots showing persona management. Panel A shows the repository dashboard for excalidraw/excalidraw with three navigation cards (View Personas, Browse Issues, Analytics), counts of 5 personas and 10 issues, a completed processing status bar, and counts of 5 personas created and 10 issues mapped. Panel B shows the Personas page with a toolbar (circled 1) containing Create Custom Persona, Generate AI Personas, and Merge Personas buttons, a search bar, and an issue mapping status panel (circled 2) showing 10 total issues, 0 unmapped, and 10 mapped. Below are four persona cards (circled 3) for Sarah (IT Coordinator, age 50, Washington D.C.), Ravi (Freelance Software Developer, age 38, Bangalore), Emily (High School Teacher, age 29, Toronto), and Carlos (Product Manager, age 42, Madrid), each showing a headshot, quote, background, goals and motivations, pain points, and Edit/Delete buttons (circled 4). Each card displays an ``AI'' tag. Panel C shows the Edit Persona dialog with two tabs: Basic Info (circled 5) containing fields for persona name, age, occupation, location, quote, tagline, and background, plus avatar options for AI-generated headshot or avatar style; and Goals and Pain Points (circled 6) showing editable lists of goals and pain points with add and remove controls. Panel D shows the Analytics Dashboard with four summary metric cards (circled 7) displaying Total Issues (10), Active Personas (5), Coverage Rate (100\%), and Repository Stars (114,473), followed by two charts (circled 8): a Label Distribution pie chart and a Persona Coverage bar chart showing issues mapped to each persona.}
\label{fig:workflow_persona_management}
\end{figure*}

\begin{figure*}[t]
\centering
\includegraphics[width=0.89\textwidth]{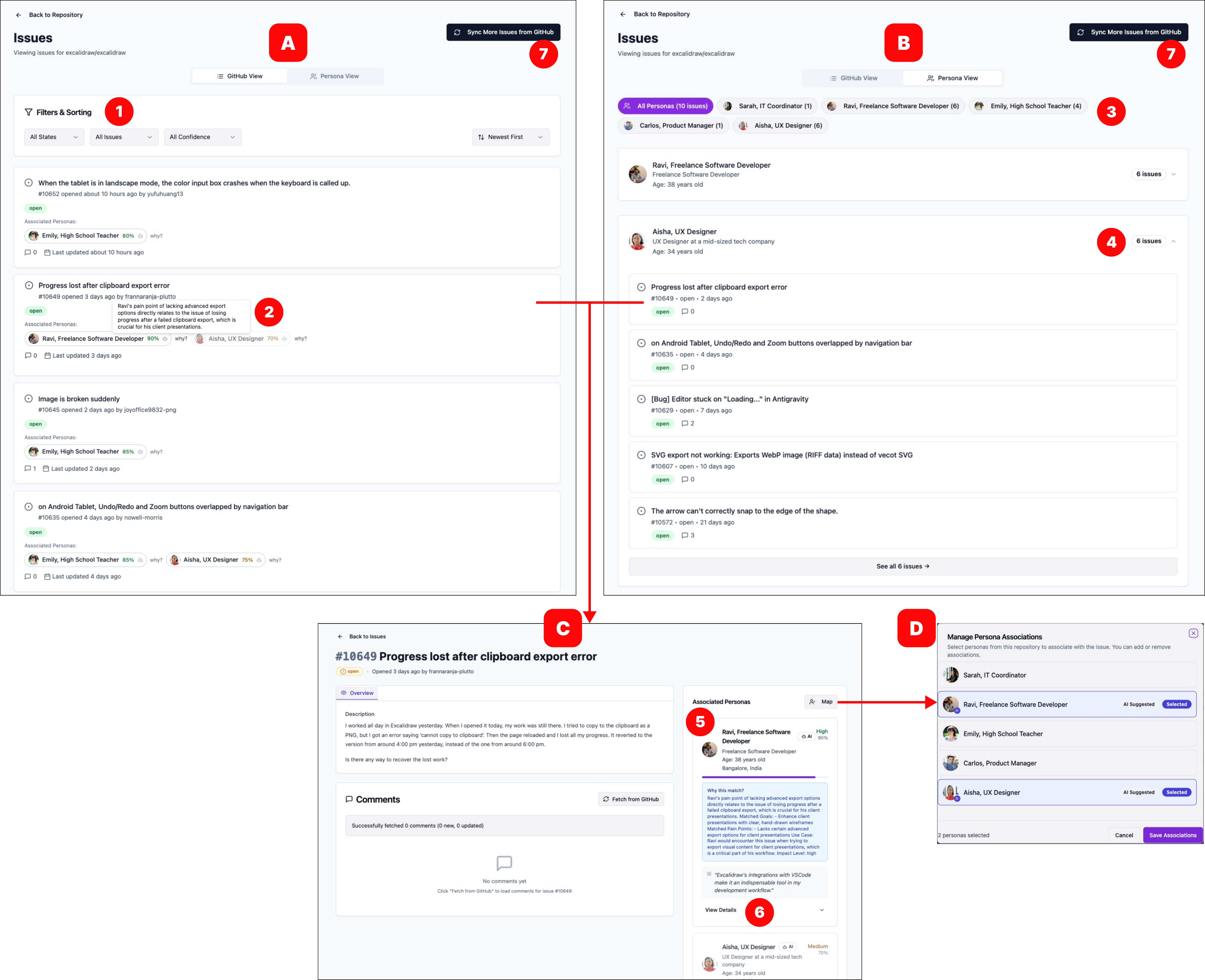}
\caption{Issue browsing and management workflow. A toggle switches between GitHub-style list view (A) and persona-grouped view (B); each provides filtering options. Issue detail page (C) shows full persona cards with confidence scores, mapping reasoning, and expandable details. An associations dialog (D) is provided for manual addition or removal of persona mappings.}
\Description{Four connected screenshots showing issue browsing. Panel A shows the GitHub View with a toggle between GitHub View and Persona View, filter controls (circled 1) for state, issue type, and confidence level, and a chronological list of issues. Each issue card shows its title, open/closed status, and associated persona badges with confidence percentages and ``why?'' buttons (circled 2). For example, ``Progress lost after clipboard export error'' is associated with Ravi (Freelance Software Developer, 90\%) and Aisha (UX Designer, 75\%). Panel B shows the Persona View with persona filter chips at the top (circled 3) for Sarah, Ravi, Emily, Carlos, and Aisha, each showing issue counts. Issues are grouped under each persona (circled 4); for example, Ravi has 6 issues and Aisha has 6 issues listed beneath their names. Panel C shows the issue detail page for issue \#10649 ``Progress lost after clipboard export error,'' displaying the full issue description on the left and Associated Personas on the right (circled 5) with a Map button. The first persona card shows Ravi (Freelance Software Developer) with a ``High'' confidence score, a highlighted ``Why this match?'' section explaining the AI's reasoning, and a representative quote. A second persona card for Aisha (UX Designer) with ``Medium'' confidence is shown below with a ``View Details'' expandable section (circled 6). Panel D shows the Manage Persona Associations dialog listing all repository personas with checkboxes; Ravi and Aisha are marked as ``AI Suggested'' and ``Selected,'' with Cancel and Save Associations buttons. A ``Sync More Issues from GitHub'' button (circled 7) appears at the top of both views.}
\label{fig:workflow_issue_classification}
\end{figure*}

\subsection{User Interaction Design}
\label{sec:features}

The final interface of PersonaFlow is organized around three workflows associated with how personas can be created, integrated, and used. Each workflow embodies our design goals while navigating the tensions and resolutions discussed above.

\subsubsection{Workflow 1: Repository Analysis and Persona Generation}
\label{sec:workflow_persona_generation}

This workflow (Figure~\ref{fig:workflow_persona_generation}) transforms a GitHub repository into data-driven personas with minimal setup (DG2). The central design challenge is the automation-control tension: generation must be effortless enough to lower adoption barriers (DG2) while giving developers meaningful agency over the output (DG4).

The process begins when users provide their GitHub repository URL (A on Figure~\ref{fig:workflow_persona_generation}). Before generation, a configuration panel (B on Figure~\ref{fig:workflow_persona_generation}) allows users to specify the number of personas (1--10), provide external documentation URLs for additional context, and add custom descriptions of their user base. These options support flexibility (DG3), allowing different projects to satisfy their unique needs.

Clicking \textit{Start Persona Analysis} initiates asynchronous processing, with a progress view (C on Figure~\ref{fig:workflow_persona_generation}) displaying stage-by-stage updates: fetching repository data, crawling external sources, analyzing domain, and generating personas. This transparency (DG4) directly addresses the preliminary user feedback about uncertainty during long-running operations.

Upon completion, a preview page (D on Figure~\ref{fig:workflow_persona_generation}) presents all generated personas for user inspection. This is where the ``generate-then-refine'' resolution (DG4) becomes concrete: users can edit any persona's details, delete personas that they feel irrelevant (\textcircled{1} on Figure~\ref{fig:workflow_persona_generation}), merge multiple personas together (\textcircled{2} on Figure~\ref{fig:workflow_persona_generation}), and manually add complementary personas or regenerate all personas altogether  (\textcircled{3} on Figure~\ref{fig:workflow_persona_generation}). After this review, users can click \textit{Save Personas}  (\textcircled{4} on Figure~\ref{fig:workflow_persona_generation}), at which point the system stores the personas and automatically begins mapping repository issues in the background (DG2).

\subsubsection{Workflow 2: Persona Management and Analytics}
\label{sec:workflow_persona_management}

This workflow (Figure~\ref{fig:workflow_persona_management}) enables the developers to inspect the personas, modify them, and recognize patterns related to user impact (DG1). The repository dashboard (A on Figure~\ref{fig:workflow_persona_management}) serves as the central hub after persona generation completes. There, users are provided with quick access to three functions: (1) View Personas, (2) Browse Issues, and (3) Analytics, while summary statistics display persona and issue counts alongside processing status (DG1). The ``\textit{Browse Issues}'' function contributes to the Issue Browsing and Management workflow (Workflow 3) and will be explained in Section~\ref{sec:workflow_issue_classification}.

The ``\textit{View Personas}'' function (B on Figure~\ref{fig:workflow_persona_management}) presents all personas in a searchable grid with a management toolbar (\textcircled{1} on Figure~\ref{fig:workflow_persona_management}) offering three actions: (1) creating custom personas manually, (2) generating additional AI personas to fill potential coverage gaps, and (3) merging two or more personas (DG3, DG4). An issue mapping status panel (\textcircled{2} on Figure~\ref{fig:workflow_persona_management}) also shows total, unmapped, and mapped issue counts, helping teams identify coverage gaps.

Each persona card (\textcircled{3} on Figure~\ref{fig:workflow_persona_management}) displays its complete profile: avatar, demographics, representative quote, background narrative, goals and motivations, and pain points and frustrations. A tag is presented at the top-right corner of each persona card to indicate whether it is generated by AI or created manually by a user (DG4). The card footer (\textcircled{4} on Figure~\ref{fig:workflow_persona_management}) also provides Edit and Delete actions for ongoing refinement (DG4). The edit dialog (C on Figure~\ref{fig:workflow_persona_management}) uses a tabbed interface to organize persona attributes. The Basic Info tab (\textcircled{5} on Figure~\ref{fig:workflow_persona_management}) allows modification of demographics, quote, tagline, and background narrative, with options to choose between AI-generated headshots or stylized avatars. The Goals \& Pain Points tab (\textcircled{6} on Figure~\ref{fig:workflow_persona_management}) presents editable lists where users can add, modify, or remove individual goals and pain points.

The ``\textit{Analytics}'' function (D on Figure~\ref{fig:workflow_persona_management}) provides aggregate insights through summary metrics (\textcircled{7} on Figure~\ref{fig:workflow_persona_management}, including total issues, active personas, coverage rate, and repository statistics) alongside two visualizations (\textcircled{8} on Figure~\ref{fig:workflow_persona_management}). The label distribution chart reveals issue categorization patterns. The persona coverage chart shows how issues are distributed across personas (DG1). These analytics aim to help teams identify blind spots and inform prioritization decisions.

\subsubsection{Workflow 3: Issue Browsing and Management}
\label{sec:workflow_issue_classification}

After personas are saved, the system maps repository issues to relevant personas. This workflow (Figure~\ref{fig:workflow_issue_classification}) is where persona insights become actionable, for developers to understand the impact of each issue on their potential users (DG1, DG3).

After clicking on ``\textit{Browse Issues}'' on the repository dashboard (A on Figure~\ref{fig:workflow_persona_management}), users will be led to the Issues interface in which the original issue discussions on GitHub are imported. The interface offers two complementary views via a toggle. The \textit{GitHub View} (A on Figure~\ref{fig:workflow_issue_classification}) preserves a familiar issue layout for developers (DG3)---issues appear chronologically, but now with persona badges indicating affected users on individual issues. Filter and sorting controls (\textcircled{1} on Figure~\ref{fig:workflow_issue_classification}) allow developers narrow results by issue state (open or closed), recency, or persona mapping confidence level. On issue cards, each persona badge includes a ``why?'' button (\textcircled{2} on Figure~\ref{fig:workflow_issue_classification}) that reveals a popover with the AI's mapping rationale, allowing a quick glance without leaving the list view of issues (DG4).

In comparison, the \textit{Persona View} (B on Figure~\ref{fig:workflow_issue_classification}) organizes the issues by the personas that they affect. Persona filter chips at the top (\textcircled{3} on Figure~\ref{fig:workflow_issue_classification}) allow viewing specific personas, with issues clustered beneath each affected persona (\textcircled{4} on Figure~\ref{fig:workflow_issue_classification}) to reveal which users face the most friction by what issues (DG1). We retained both views because participants in the preliminary user study had different preferences; some wanted personas layered onto existing mental models, while others valued the perspective shift of persona-first organization.

By clicking on an issue card on either view, the user will be led to the issue detail page (C on Figure~\ref{fig:workflow_issue_classification}). This page displays full persona cards on the side of the issue discussion details (DG3). The default view of each persona card  (\textcircled{5} on Figure~\ref{fig:workflow_issue_classification}) shows its role and basic demographic information, together with a confidence score (0--100\%) of mapping this persona to the issue, AI generated reasoning for why this persona would care about the issue (DG4), and representative quotes from persona profiles. Users can expand each persona card (\textcircled{6} on Figure~\ref{fig:workflow_issue_classification}) to view its details, such as goals and pain points. When associations feel incorrect, users can click the ``Map'' button (D on Figure~\ref{fig:workflow_issue_classification}) to open the persona associations dialog, where they can manually add or remove personas to be associated with the issue; AI-suggested associations are clearly labeled on this dialog to provide transparency (DG4).

Finally, on both views, users can re-sync issues with GitHub (\textcircled{7} on Figure~\ref{fig:workflow_issue_classification}) as repositories evolve, with options to fetch all new issues or issues with specific IDs, labels, or date ranges (DG3). Newly imported or updated issues are automatically mapped to existing personas, keeping the persona association current without manual effort (DG2).

\subsection{Implementation}

\subsubsection{System Architecture}

\begin{figure*}[t]
\centering
\includegraphics[width=\textwidth]{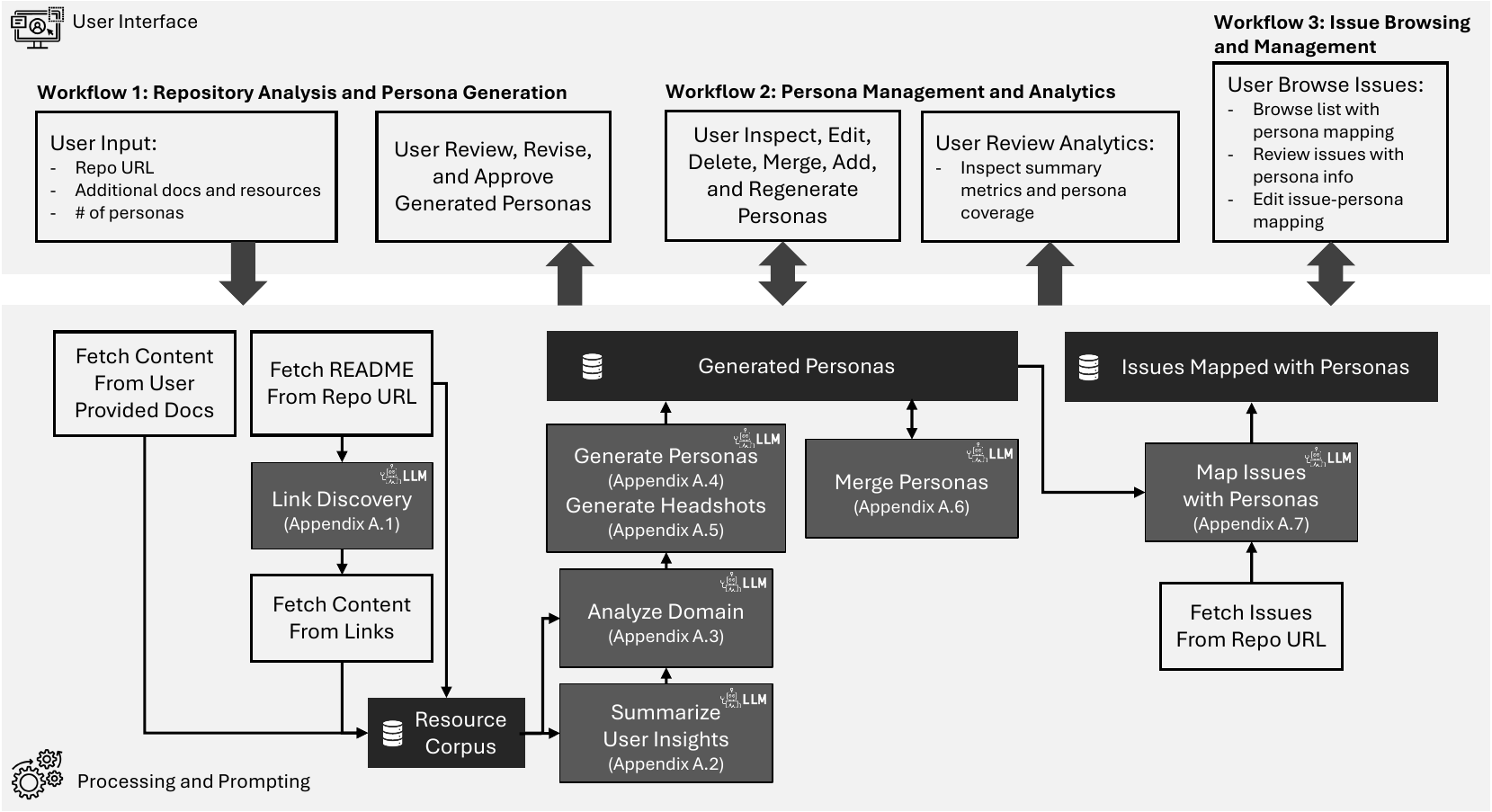}
\caption{System architecture of PersonaFlow, illustrating how the three user workflows connect to backend processing components. Shaded boxes with ``LLM'' labels are components that used an LLM (GPT-4o and Gemini 1.5-Flash in our case); the specific prompts are provided in the appendix sections indicated in the boxes. Black boxes with a data logo are storage components in the database.}
\Description{A two-layer architecture diagram. The top layer labeled ``User Interface'' shows three workflows as white boxes with black borders: Workflow 1 (Repository Analysis and Persona Generation) contains ``User Input: Repo URL, Additional docs and resources, number of personas'' and ``User Review, Revise, and Approve Generated Personas''; Workflow 2 (Persona Management and Analytics) contains ``User Inspect, Edit, Delete, Merge, Add, and Regenerate Personas'' and ``User Review Analytics: Inspect summary metrics and persona coverage''; Workflow 3 (Issue Browsing and Management) contains ``User Browse Issues: Browse list with persona mapping, Review issues with persona info, Edit issue-persona mapping.'' The bottom layer labeled ``Processing and Prompting'' shows the data flow: User-provided docs and the README fetched from the repo URL feed into a Link Discovery LLM step (Appendix A.1), which fetches content from discovered links. All content flows into a Resource Corpus database. From there, two parallel LLM steps process the corpus: Summarize User Insights (Appendix A.2) and Analyze Domain (Appendix A.3). Their outputs feed into Generate Personas (Appendix A.4) and Generate Headshots (Appendix A.5), producing a Generated Personas database. A Merge Personas LLM step (Appendix A.6) also connects to this database. Issues fetched from the repo URL and the Generated Personas feed into a Map Issues with Personas LLM step (Appendix A.7), producing an Issues Mapped with Personas database. Arrows connect the databases back up to the corresponding user interface workflows.}
\label{fig:system_architecture}
\end{figure*}

PersonaFlow is implemented as a web application with a React frontend and a Django backend, using PostgreSQL for data persistence. Figure~\ref{fig:system_architecture} illustrates the system architecture showing how the three user workflows connect to backend processing components. The system connects with several external services: the GitHub REST API for repository metadata and issue data, the OpenAI API for AI-powered persona generation and issue mapping, and the Google Gemini API for avatar image generation. Once users initiate persona generation, the backend fetches repository data from GitHub and constructs appropriate prompts to be sent to the LLM APIs. The LLM outputs are parsed, validated, and stored in the database. After users inspect and save personas, the backend automatically initiates issue synchronization and mapping.

Because persona generation involves multiple time-consuming steps (content fetching, AI analysis, avatar generation) that can take several minutes to complete, we used an asynchronous task processing mechanism with Celery and Redis as the message broker. When users initiate long-running operations, the backend immediately returns a task identifier while processing continues. The frontend polls a dedicated status endpoint to retrieve real-time progress updates and displays stage-by-stage feedback to users. This approach ensures a responsive experience despite computationally expensive AI operations.

\subsubsection{LLM Prompt Design}
\label{sec:impl_prompts}

As shown in Figure~\ref{fig:system_architecture}, we included three main prompting pipelines in PersonaFlow.

\textbf{Persona Generation Pipeline.} We use a multi-stage approach in generating the personas. First, we used a \textit{Link Discovery Prompt} (see Appendix~\ref{sec:app_link_discovery}) to find links in the project README file that could help understand the end users. The content fetched from those links, together with other user-provided resources, is used to build a resource corpus. Then, the \textit{User Insights Summarize Prompt} (see Appendix~\ref{sec:app_user_insights}) and the \textit{Domain Analysis Prompt} (see Appendix~\ref{sec:app_domain_analysis}) are used to extract structured user-related and domain-related context from this resource corpus. Finally, the \textit{Personas Generation Prompt} (see Appendix~\ref{sec:app_persona_generation}) creates personas based on information extracted from the prior steps; this prompt emphasizes diversity across user characteristics and usage contexts, as well as adherence to the user and domain insights. Overall, this chained, multi-stage approach aims to ensure that personas are grounded in actual repository data rather than generic assumptions about software users.

\textbf{Persona Merging Pipeline.} In both Workflow 1 and Workflow 2, users can merge two or more personas when they identify overlaps. This is achieved through the \textit{Persona Merge Prompt} (see Appendix~\ref{sec:app_merge}). This prompt combines input personas, while aiming to preserve a coherent narrative while representing the common characteristics of the original personas.

\textbf{Issue Mapping Pipeline.} The \textit{Issue-Persona Mapping Prompt} (see Appendix~\ref{sec:app_batch_matching}) connects GitHub issues to relevant personas. Rather than keyword matching, it asks whether a persona would plausibly write or care about each issue given their goals and workflows. Anti-pattern checks are included to reject superficial matches (e.g., generic reasons that resolving issues can benefit ``all users''). To support developers in verifying and overriding suggestions, the prompt also outputs confidence scores with reasoning.

\vspace{6pt}
\textbf{\textit{API Cost Estimation}.} Generating personas from a repository requires four text-based LLM calls (i.e., link discovery, user insights summarization, domain analysis, and persona generation, using GPT-4o) plus one image generation call per persona for avatar creation (using Gemini 1.5-Flash). Each issue mapping requires one LLM call. In our user study, the total cost per repository (each with four generated personas mapped with 20 issues, see Section~\ref{sec:methods_data}) ranged from \$0.50~CAD to \$1.00~CAD, with issue mapping accounting for the majority of the cost (\$0.40~CAD to \$0.70~CAD).

\subsubsection{Assessing Persona Generation and Issue Mapping Quality}
While it is challenging to systematically and quantitatively evaluate the quality of the generated personas and the accuracy of issue-persona mappings, we performed frequent manual inspections during our prompting and tool development processes. Here, we present example personas generated by PersonaFlow across a diverse set of OSS projects---from creative tools to developer infrastructure---showing how the system adapts to different product domains (see Table~\ref{tab:persona_examples}). Each persona was automatically created from a single input of the repository URL, without other manual input, representing a baseline for users of PersonaFlow to work on. The examples illustrate how goals and pain points reference specific product features rather than generic professional aspirations or user tendencies. The issue mapping example of each persona was randomly selected from the ten most recent issues mapped to the persona. The issue mapping is further explained by how the issue affects the persona's goals and pain points, which is indicated in the AI's reasoning text (see the \textit{Issue Mapping} parts on Table~\ref{tab:persona_examples}).

\begin{table*}[t]
\centering
\caption{Examples of AI-generated personas from repositories. Each persona was automatically created from a single input of the repository ID, without other manual input. For each persona, one example issue mapped to the persona is provided, including the issue's ID and title, our manual summary of the issue content, and the AI's confidence score and reasoning.}
\label{tab:persona_examples}
\scriptsize
\begin{tabular}{p{0.11\textwidth}@{}p{0.87\textwidth}}
\toprule
\multicolumn{2}{@{}l}{\textbf{Repository: musescore/MuseScore} \textit{(Open-source music notation software)}} \\
\midrule

\textbf{Persona} &
\textbf{Yuki Tanaka}, 32, Tokyo, Japan --- Freelance Composer and Arranger \\[0.15em]

\multicolumn{1}{r}{\textit{Quote}} &
\textit{``I need my scores to look professional, but I also need to create them quickly to meet tight deadlines.''} \\[0.15em]

\multicolumn{1}{r}{\textit{Background}} &
Yuki has been composing music since her teenage years, inspired by both traditional and modern Japanese music. Her freelance work ranges from creating scores for indie films to arranging music for local orchestras. Her clients demand high-quality scores under tight deadlines. \\[0.15em]

\multicolumn{1}{r}{\textit{Goals}} &
$\bullet$ Complete initial score drafts in half the usual time using MuseScore's easy note entry \newline
$\bullet$ Transform rough musical ideas into polished scores within a week for client rehearsals \newline
$\bullet$ Develop professional presentation skills for quick adjustments to scores \\[0.15em]

\multicolumn{1}{r}{\textit{Pain Points}} &
$\bullet$ Frequent struggle with aligning musical elements neatly, taking unnecessary hours due to limited spacing options \newline
$\bullet$ Manual note entry requires extra effort when intuitive shortcuts could streamline workflow \newline
$\bullet$ 10\% of scores lose formatting fidelity due to incompatible export settings, forcing last-minute adjustments \\[0.15em]

\textbf{Issue Mapping} &
\textit{\#28950 --- Dynamics popup should not cover the staff to which the dynamic is applied} \newline
\textit{Summary:} When editing dynamics placed above a staff, the popup appears below and obscures the notation. \newline
\textit{AI Reasoning:} ``This UI issue directly affects Yuki's goal of creating professional-looking scores efficiently. The popup obstruction would slow her workflow during time-sensitive client projects.''\\

\midrule
\multicolumn{2}{@{}l}{\textbf{Repository: flameshot-org/flameshot} \textit{(Cross-platform screenshot software)}} \\
\midrule

\textbf{Persona} &
\textbf{Nina Patel}, 34, Mumbai, India --- Marketing Manager at a Tech Startup \\[0.15em]

\multicolumn{1}{r}{\textit{Quote}} &
\textit{``I need tools that keep up with my pace and let me share ideas instantly.''} \\[0.15em]

\multicolumn{1}{r}{\textit{Background}} &
Nina has been in the tech marketing field for over a decade, transitioning from traditional media to digital strategies. With Mumbai's vibrant startup ecosystem, she often needs to capture and annotate screenshots for presentations and online campaigns. \\[0.15em]

\multicolumn{1}{r}{\textit{Goals}} &
$\bullet$ Achieve polished presentations with annotated visuals using integrated editing tools \newline
$\bullet$ Transform raw screenshots into shareable content within 15 minutes for quick team feedback \newline
$\bullet$ Customize appearance to match brand guidelines for consistent visual communication \\[0.15em]

\multicolumn{1}{r}{\textit{Pain Points}} &
$\bullet$ Weekly struggle with editing screenshots taking 30+ minutes due to switching between apps \newline
$\bullet$ Setting up tools requires more time than ideal, blocking swift adaptation to new platforms \newline
$\bullet$ 20\% of shared visuals are delayed due to inconsistent formatting, forcing last-minute adjustments \\[0.15em]

\textbf{Issue Mapping} &
\textit{\#4062 -- Text tool is ignoring first edit}\newline
\textit{Summary:} First text edit reverts to original content; subsequent edits work correctly. \newline
\textit{AI Reasoning:} ``Nina's need for rapid sharing and customization of visuals aligns directly with this text editing bug. The issue would cause frustration during time-sensitive campaign work where quick annotations are essential.''\\

\midrule
\multicolumn{2}{@{}l}{\textbf{Repository: calcom/cal.com} \textit{(Open-source scheduling infrastructure)}} \\
\midrule

\textbf{Persona} &
\textbf{Marta Gonzalez}, 45, Madrid, Spain --- Operations Manager at a Small Consultancy \\[0.15em]

\multicolumn{1}{r}{\textit{Quote}} &
\textit{``Having a reliable scheduling system that clients can access without issues is crucial for our operations.''} \\[0.15em]

\multicolumn{1}{r}{\textit{Background}} &
Marta has been managing operations for small consultancies for over 15 years. She's responsible for ensuring smooth client interactions and internal coordination, prioritizing tools that are reliable and easy for clients to use. \\[0.15em]

\multicolumn{1}{r}{\textit{Goals}} &
$\bullet$ Streamline client booking processes with a branded scheduling interface \newline
$\bullet$ Enhance client satisfaction by reducing booking errors \newline
$\bullet$ Integrate Cal.com with existing CRM systems for unified workflow \\[0.15em]

\multicolumn{1}{r}{\textit{Pain Points}} &
$\bullet$ Occasional glitches in client-facing booking pages affect professional image \newline
$\bullet$ Limited customization in branded interface without technical input \newline
$\bullet$ Security concerns when integrating with CRM systems \\[0.15em]

\textbf{Issue Mapping} &
\textit{\#23463 -- Slot Availability Bug: Extra unwanted slots appear despite 2-hour blocks}\newline
\textit{Summary:} In medical scheduling, extra time slots appear despite configuring 2-hour blocks, allowing patients to select unintended appointment times.  \newline
\textit{AI Reasoning:} ``This directly affects Marta's goal of reducing booking errors. Unintended slots in a client-facing system could cause scheduling confusion and undermine trust in her consultancy's professional image.''\\

\midrule
\multicolumn{2}{@{}l}{\textbf{Repository: pocketbase/pocketbase} \textit{(Open-source backend in a single file)}} \\
\midrule

\textbf{Persona} &
\textbf{Olivia Martinez}, 45, San Francisco, USA --- IT Manager at a Non-Profit Organization \\[0.15em]

\multicolumn{1}{r}{\textit{Quote}} &
\textit{``We need a cost-effective backend solution that even our non-technical staff can manage.''} \\[0.15em]

\multicolumn{1}{r}{\textit{Background}} &
Olivia manages a small IT team responsible for supporting the technological needs of a non-profit. With limited resources, she prioritizes solutions that are easy to manage and don't require extensive technical expertise. Her team often includes volunteers who assist with tech tasks. \\[0.15em]

\multicolumn{1}{r}{\textit{Goals}} &
$\bullet$ Implement a secure, low-cost backend solution using PocketBase \newline
$\bullet$ Empower non-technical staff to manage app content via the Admin Dashboard \newline
$\bullet$ Reduce IT resource strain by leveraging built-in user management \\[0.15em]

\multicolumn{1}{r}{\textit{Pain Points}} &
$\bullet$ Difficulty training non-technical staff due to lack of intuitive setup guides \newline
$\bullet$ Missing audit log features complicate compliance reporting \newline
$\bullet$ Potential downtime during updates due to lack of backward compatibility \\[0.15em]

\textbf{Issue Mapping} &
\textit{\#5096 -- Proposal: Hide record create and edit controls} \newline
\textit{Summary:} Feature request to limit UI controls for non-admin users to prevent accidental modifications. \newline
\textit{AI Reasoning:} ``This would help Olivia empower non-technical staff by preventing accidental data modifications---a key concern for teams with mixed technical abilities.''\\

\bottomrule
\end{tabular}
\end{table*}

\section{User Study Methods}
\label{sec:study_methods}
We conducted a user study using PersonaFlow as a probe to understand how OSS developers may engage with tools that embed AI-generated personas in the context of issue handling. 
Specifically, we explored (1) developers' pre-existing empathic practices when responding to issue reports, (2) how persona exposure influences their mental models and communication with users, and (3) their perceptions of the various features of PersonaFlow for user-centered development. The study is approved by the ethics review board of the involved institutions.

\subsection{Participants}
Thirteen software developers who actively contribute to OSS projects were recruited through purposive sampling via developer communities, OSS mailing lists, LinkedIn, and Upwork. Participants represented diverse backgrounds in terms of the primary programming languages they use, OSS experience, age ranges, and geographic regions (spanning North America, Central America, Europe, the Middle East, and Asia). The sample was predominantly male (12 male, 1 female); while this is unfortunately imbalanced, it represents the general gender distribution among OSS contributors~\cite{Zhao2023OSSGender}. Participants' experiences of OSS contribution ranged from beginners who just started to get involved in OSS projects to veterans who contributed to OSS for more than 20 years. The participants' characteristics are detailed in Table~\ref{tab:participant-info}.

\begin{table*}[t]
\centering
\small
\caption{Summary of user study participants. Participants work on diverse OSS domains, with experience ranging from newcomers ($<$1 year) to veterans (20+ years).}
\label{tab:participant-info}
\begin{tabular}{llllllcc}
\toprule
\textbf{ID} & \textbf{Profession} & \textbf{Primary Language} & \textbf{OSS Domain} & \textbf{Country} & \textbf{Age Range} & \textbf{Gender} & \textbf{OSS Exp.} \\
\midrule
P1  & PhD Student           & Python     & Data Science        & Canada    & 30--34 & F & 1-3 years  \\
P2  & Software Developer    & PHP        & Web/CMS             & Nepal     & 35--40 & M & 1-3 years  \\
P3  & Software Developer    & TypeScript & Productivity        & Canada    & 25--29 & M & 1-3 years   \\
P4  & Freelancer            & C++        & Multimedia          & Poland    & 25--29 & M & $<$1 year  \\
P5  & AI Engineer           & Python     & Developer Tools     & Canada    & 25--29 & M & 5-10 years  \\
P6  & Author                & PHP        & Messaging           & USA       & 40+    & M & $>$20 years \\
P7  & Sr. Software Engineer & Python     & Infrastructure      & USA       & 40+    & M & 10-20 years \\
P8  & Developer Advocate    & Python     & Mobile Database     & Taiwan    & 40+    & M & 10-20 years \\
P9  & Student               & JavaScript & Developer Tools     & Nicaragua & 20--24 & M & 1-3 years   \\
P10 & Student               & TypeScript & Developer Tools     & Canada    & 20--24 & M & $<$1 year  \\
P11 & Freelancer            & TypeScript & Video/Multimedia    & Nicaragua & 20--24 & M & $<$1 year  \\
P12 & Consultant            & C          & Document Processing & USA       & 40+    & M & $>$20 years \\
P13 & Software Engineer     & C++        & Game Engine         & Egypt     & 25--29 & M & $<$1 year  \\
\bottomrule
\end{tabular}
\end{table*}

\subsection{Data Preparation}
\label{sec:methods_data}
Since PersonaFlow requires processing time (typically five minutes) to generate personas and classify issues, we asked participants during recruitment to provide two GitHub repositories they actively contribute to or are familiar with so that we can prepare the data ahead of the study sessions. Prior to the study, we ran PersonaFlow on each participant's selected repository to generate four personas for each repository. Additionally, we used the tool to retrieve and map the 20 most recent open issues to the generated personas. No editing to the personas or the issue mapping was performed to ensure that the participants experience the original output of the tool during the study, including potential AI errors and discrepancies. Subsequently, we selected one issue per repository from the issues retrieved by the tool, to prepare for the user study tasks described in the next section. Because we want to understand how PersonaFlow can affect participants' user-centeredness tendency, we selected the issue iteratively with the following process: we first identified issues less than one year old in which the participant did not involve with, to ensure that they carry out a fresh engagement with the issues during our study; we then filtered out issues that were overly technical or purely code-oriented (e.g., bug reports posted by someone with strong technical background); from the remaining, we then selected an issue where user frustration or contextual ambiguity was the most evident. 

\subsection{User Study Process}
All study sessions were conducted remotely via Zoom. With participants' consent, sessions were video recorded with screen capture. Each session lasted approximately 90 minutes and contained the following parts.

\textit{Background Interview.}
We began each session with background questions about their professional role, programming experience, and OSS contribution history. We then explored their existing empathic practices by asking how they typically handle issue reports and whether they had prior experience with personas.

\textit{PersonaFlow's Impact on Issue Response.}
Subsequently, each participant completed the following three tasks using one of the repositories they had identified during recruitment:

First, participants reviewed the issue we selected from this repository using PersonaFlow's interface with persona information hidden. Without any additional context about the user, they described their initial assumptions about the user who submitted the issue, a priority rating on a scale of 1 (Low Priority) to 10 (Critical Priority), and a written response they would post to the user. This task captured their natural approach to issue triage before any intervention.

Second, participants watched a brief introduction video to personas created by the NNGroup\footnote{\url{https://www.youtube.com/watch?v=rv9yfrV-EAs}}. They then were asked to explore at least two of the AI-generated personas we had prepared for the repository on PersonaFlow. They were asked to discuss what stood out about each persona to them and to propose a feature or improvement that would address one persona's specific goals and frustrations. This task allowed participants to engage with the persona concept and our tool before applying it to a real issue.

Finally, participants returned to the original issue from the first task, using the complete version of PersonaFlow. Now supported by the personas PersonaFlow had created and automatically matched to the issue, as well as other features of the tool, they rewrote their response to the issue post. Immediately after, we showed them their original and revised responses side-by-side and asked them to reflect on differences in language, tone, and content. We also asked whether their priority assessment of the issue had changed. This comparison allowed us to observe how persona exposure influenced their communication approach.

\textit{Free Exploration of PersonaFlow's Features.}
Participants were then asked to freely explore the tool for about 30 minutes. They could either continue with the same repository or analyze the other repository they had provided during recruitment. During this exploration, we asked questions about their feedback on various tool features, as well as their overall impressions of the tool's potential for supporting user awareness and understanding during OSS development practices.

\subsection{Data Analysis}
The recordings of the study sessions were fully transcribed. We conducted a thematic analysis~\cite{Vaismoradi2013, aronson_pragmatic_1995} to identify patterns and insights within the data. The first author led the data analysis process, supported by frequent meetings and discussions with other authors for the synthesis of themes and clarification of any ambiguity. Specifically, we first assigned structural codes to different segments of the transcripts, aligning these codes with the overall structure of our study to facilitate subsequent analysis. Next, an inductive coding approach was employed to identify themes and concepts from participants' comments and feedback. Descriptive codes were generated initially from the raw data, capturing distinct ideas, opinions, or experiences expressed by participants. The codes were then iteratively grouped into categories, which were further organized into themes. The themes were then arranged into four overall groups: (1) participants' current issue handling approach, (2) the impacts of personas incorporated in PersonaFlow on participants' user empathy, (3) participants' perceptions and feedback on PersonaFlow, and (4) factors that can affect the adoption of tools like PersonaFlow in practice. The coding and grouping were done collaboratively through multiple rounds of discussions among the authors.

\section{User Study Results}
Our analysis shows that personas and the various features of PersonaFlow have affected how participants perceive and address users' questions and requests raised in GitHub issues. Participants also provided feedback on our tool and discussed factors that can influence its adoption in their real-world practices. We present these findings in the following sections. In addition, Appendix~\ref{sec:appendix_case_studies} presents two detailed case studies from our study, walking through the full workflow of two participants---from repository input and system processing to persona generation, issue mapping, and developer interaction---on two real repositories. 

\subsection{The Pre-Existing Technical Orientation in Issue Handling}

Before exposure to PersonaFlow, developers exhibited patterns consistent with the empathy gap identified in prior research on distributed software development~\cite{gunatilake2024enablers, carmel2001work}. While some participants reported attempting empathetic behaviors (e.g., P10 described trying to \textit{``put myself in [the] position of someone without technical prowess''} when testing UI), these efforts were inconsistent and unsystematic. We identified the following common tendencies of issue handling from participants' comments and behaviors.

When handling issues, \textbf{developers lacked a user-centered perspective, prioritizing technical validation over understanding user needs}. As P12 explained: \textit{``It's about effective communication and reduction of time spent on a particular issue,''} framing developer-user interaction as an efficiency problem rather than a human connection. Similarly, P4 noted their focus on \textit{``solutions... about the technicalities,''} treating issues as puzzles to solve rather than people to help. This technical orientation extended to strategies to respond to issue requests: P10 described a systematic workflow of \textit{``try and get them [users] to give me step by step what they did... reproduce it on my end... do a lot of logs.''} While methodologically sound, this approach treats issue reporters primarily as information sources for software problem diagnosis, where user goals, context, and frustration are rarely factored into discussions.

Additionally, \textbf{developers struggled to infer user characteristics and emotions from written reports}. To assess the expertise of the software users, they relied on existing artifacts as proxies. For example, P8 observed how writing style signaled competence: \textit{``It's properly written... [probably by] someone who has written an issue before... So it sounds like a person actually has a clue of what they're doing.''} P9 described checking GitHub profiles: \textit{``If it's a new account... they don't know much about coding, they just create the GitHub account and put the issue.''} While these heuristics helped prioritize, they risked dismissing legitimate issues from less technical and novice users. Detecting emotions proved even harder---frustration was only visible through explicit signals, which are also sparse. P3 noted: \textit{``It would be difficult unless they make it very clear, like they're typing in uppercase and they're swearing... But if they're just typing normally, I would assume they're neutral.''}

Beyond technical orientation, \textbf{developers faced systemic barriers that actively inhibited empathetic engagement}. For example, P13 discussed the constant flow of issues that OSS developers often face, which can limit the time and energy available to consider the perspectives of others: \textit{``The open-source developers are the ones that actually carry the community and no one ever thinks about them... They are always getting just issues and then issue after issue.''}  Indeed, the high workload created an emotional burden on developers that can foster defensive stances. P12 depicted this stance bluntly: \textit{``RTFM. Read the f**king manual,''}, illustrating their frustration with handling constant requests and their expected ways to resolve this tension:  users should self-educate before engaging. This expectation is reasonable for certain issues, but can create a high barrier to entry that excludes users who are unfamiliar with the code base and documentation. 

When prioritizing issues, \textbf{developers relied on aggregate metrics rather than individual user needs}. P2 described a threshold-based approach: \textit{``If any user comes and says if that's not working for them but it's working for 500 customers... we will collect feedback until a few of the customers hit that threshold.''} This meant that individual users, no matter how frustrated, were systematically deprioritized until more reports surfaced. Technical severity further shaped prioritization, with UX issues ranked below functional failures. For example, P9 dismissed a UI/UX issue as low priority because \textit{``it's not like breaking anything, it's just an extra step the user has to do.''} Such a perspective and approach often leave persistent usability issues affecting individuals unaddressed.

\subsection{Persona Impact on Developers' User Empathy}
Of the 13 participants, eight (61.5\%) modified their responses after seeing the personas, a notable shift given that participants had just written their initial response moments earlier. Among the five who did not revise, two considered their original responses already appropriate for any user type (P2, P6), two reported internal attitudinal shifts without visible textual changes (P8, P9), and one would only modify their response conditionally for a different persona type (P11). For those who changed their responses, the nature of the changes revealed empathy activation: responses shifted from purely technical to acknowledging user experience. For example, P1's initial response advised users to ``avoid the extra clicking,'' implicitly blaming user behavior; but after seeing the persona, P1 added: \textit{``I can understand, with this bug, your workflow may be affected and you may feel frustrated.''} Similarly, P3's generic acknowledgment of issue request transformed into: \textit{``I understand that you are having trouble with this because organization truly matters to you... You are not going to have to deal with this pain anymore.''} P3 also raised their priority rating (measured on a 1--10 scale from Low to Critical) from 3 to 8, demonstrating that empathy translated into concrete prioritization changes.
The impacts of the persona also manifested in participants' verbally reported changed feelings about the users. Regarding these impacts, we have identified the following themes.

Notably, \textbf{personas had varying effects across participants}. Some participants expressed empathy shifts, although without visible behavioral changes. For example, P8 reported that personas provide \textit{``some positive impact on having a name and having a face to a person... [though] it's an AI generated person;''} but they did not perform any change to their response comment to the issue. Others demonstrated conditional calibration towards a certain group of users. For instance, after assessing one issue with the mapped persona of a senior developer, P11 judged their response to the issue appropriate, and explained that they would modify their response for a junior persona who was \textit{``first time working with Remotion.''} For many, however, exposure to the personas translated into major changes in their communication in the issue discussion thread. For instance, P5 explained how seeing the technical experience of the persona shaped their approach: \textit{``Since it's a junior developer... I will try to maybe add some links to how to install the library properly.''} Participants also added explicit personal acknowledgment of user impact when adjusting their response after seeing the personas. For example, P10 added this sentence in their response to personally address the issue poster as a user: \textit{``Hopefully this alleviates the friction for your debugging.''}

Generally speaking, \textbf{personas steer the developers' pre-existing mindset, that is primarily code-centric, closer to user-centric thinking}, disrupting their default technical orientation. P10 succinctly summarized how the persona reoriented their focus: \textit{``As soon as you have this persona that you're responding to... it helps me better ground myself in what the user needs... [without it] you're thinking just in terms of the code.''} These shifts are manifested in the following three ways. First, \textit{\textbf{the personas enabled participants to see users as concrete individuals}}, transforming abstract issue reporters into individuals with backgrounds, goals, and frustrations. For instance, P10 observed how personas humanized what had been an anonymous exchange: \textit{``It's easy to kind of say, `oh, like whatever, there's this little bug, I'll get to it when I get to it.' But then once you see these personas and you get the idea of why they're being affected by it, how it's harming their workflow, I find I am a lot more empathetic.''} Second, \textit{\textbf{the personas allowed the participants to think about a broader spectrum of users}}. For example, P13 observed: \textit{``Without these personas, probably they wouldn't think about all of these users for their engine...[such as] a professor who's trying to teach... this is eye opening to say the least.''} Finally, \textit{\textbf{the personas helped participants understand the impact of issues on the users}}. In this, P1 described an attitude shift from annoyance to understanding, when looking at a persona generated for repositories they were familiar with: \textit{``From negative to positive... this person tried to integrate my code into his workflow but it does not work well. So I may have a more positive idea about why I should update.''}

Moreover, \textbf{while some participants remained skeptical of the usefulness of personas concerning empathy building, they still suggested benefits of PersonaFlow on issue impact analysis and productivity}. For example, P10 described using persona analytics for issue prioritization: \textit{``If I see one Persona that has like eight bugs and the others have two, maybe I'll prioritize that persona [with eight bugs] so those types of users aren't getting the short end of the stick.''} Others rationalized the persona's benefits purely as efficiency gains. P9 explicitly rejected the empathy frame: \textit{``I don't feel it more as an empathy tool... I feel it more [as a] way to easily triage the issue.''} These expected usages suggest that using personas to approach the issue discussion space can nudge user-centered behavior even among developers who reject empathy as a goal: behavioral change occurred independently of conceptual buy-in. The case of P12 further demonstrated this observation: P12 stated \textit{``I don't buy the persona concept,''} but still added a practical workaround to their response after seeing the persona.

\subsection{Participants' Perceptions and Feedback to the Design of PersonaFlow}

Participants provided feedback on both the persona content and the features of PersonaFlow. Overall, they valued the simplicity of the tool. For example, P7 praised the interface: \textit{``Hugely easy to use.''} P4 also noted: \textit{``The UI of the tool is really nice... It's clear what is where and it seems straightforward to use.''} The concise persona formats also prevented information overload, as P4 mentioned: \textit{``It's good that [the personas are] short and to the point... each persona is not a three-page essay.''} Beyond general simplicity and usability, developers discussed which persona elements were effective at surfacing user context and how the tool design supported empathetic engagement.

\subsubsection{Feedback to Persona Content Elements}

Among the persona content elements, \textbf{participants valued pain points, goals, and backgrounds for surfacing user context}. For example, P6 valued how these elements synthesized user information: \textit{``I like how it ... shows you how they're involved... what their main goals are and the main competencies they have.''} P9 also appreciated: \textit{``I really like the pain points and frustrations because it let me know their goals and motivations.''} 
P10 found it useful to have these elements always available whenever a persona is referred to, including on the issues page: \textit{``Being able to click on their goals, motivations, pain points... It's very good to always keep in mind what their goals are.''} 

However, \textbf{participants held competing opinions about the demographic information of the personas}. Some viewed such information as irrelevant noise. For example, P8 dismissed those attributes: \textit{``Location. Age. Name. And the picture? None of those things [are useful]... literally the only thing I would probably find helpful is the role.''} Others valued demographics precisely because they enabled human connection. For instance, P10 noted that \textit{``having the pictures too... very like humanizes the personas,''} while P4 saw the primary value as \textit{``a nice tool to remind developers that there is a person on the other side.''}

\subsubsection{Participants' Strategies Navigating the Automatically Generated Content}
Developers did not passively accept AI-generated personas. Instead, they drew on their domain expertise and existing knowledge of users to critically evaluate whether each persona accurately represented their user base.

To calibrate trust, \textbf{developers validated AI-generated personas by testing them against their existing knowledge of users}. For example, P4 trusted a persona because it matched real people: \textit{``There are people who actually I know and they produce music with the help of Audacity... I am not surprised this persona appears here.''} P8 also validated a persona when goals mirrored the product's core value: \textit{``Being offline capable was the number one argument for why Realm is so amazing. So this persona has exactly the goals that make sense.''}

Conversely, \textbf{when personas conflicted with developers' mental models, they tended to be rejected}. For instance, P5 dismissed non-technical personas for technical tools: \textit{``I don't think a project manager would be interested in knowing anything about Python dotenv... this is more popular for developers.''} P8 caught technical inconsistencies: \textit{``Realm Swift is for iOS because it's Swift. So the whole backend thing [described in this persona] I'm not really sure about.''} However, detailed backgrounds sometimes enabled reconsideration. P5 initially dismissed the project manager persona but changed their view after reading the context: \textit{``That becomes more interesting because our project manager will make sure that we're using a proper tool.''}

Despite critical evaluations, \textbf{developers accepted AI limitations while remaining vigilant about persona accuracy}. P8 articulated a pragmatic stance: \textit{``You can't trust 100\%, but you get a lot of productivity boost out of it... AI is sometimes wrong. We're humans, we're also sometimes wrong.''} However, developers recognized that persona accuracy carried significant stakes. P4 warned: \textit{``Doing personas badly, like low-quality personas, can possibly be devastating for the project... if you misrepresent the user base.''} P3 suggested that AI could help maintain objectivity: \textit{``It would be nice if the AI could keep me non-biased... detect `I think you're going out of context here.'~''}

\subsubsection{Feedback to the Tool's Features}

For large-scale projects, \textbf{automated persona generation reduced workload while preserving human control}. P8 appreciated the scale: \textit{``Getting those personas out of a repository, especially if we look at 6700 issues. You're not going to read them. You need some kind of tool to help you set up those personas.''} The confidence indicator and editing feature were valued for correcting AI outputs. P8 explained: \textit{``The 85\% confidence basically tells us it's just 85\% accurate... having an edit feature for exactly that reason I think super helpful.''} P4 described the ideal workflow: \textit{``I would probably create as many [personas] as possible... try to merge the obvious ones... edit them as I go.''} P5 also noted that merging helped make more reasonable personas: \textit{``Merging both of them to have a CTO with an interesting background in neural network development is a persona that is plausible.''}

When managing large backlogs of user-reported issues, \textbf{issue-to-persona mapping aided prioritization and decision support}. For example, P2 appreciated this integration: \textit{``Pulling the issues from GitHub... and associating them to the personas automatically... help us to categorize the issues more precisely.''} P10 found the rationale generated for persona to issue mapping useful as a quick reference: \textit{``I like this little section about why this match... if I don't really remember the full persona, at least I get a little summary of why it matches.''} Moreover, developers wanted further transparency to inform their judgment, as P1 explained: \textit{``I'm interested in why the other persona does not match... I want to have both perspectives... and make my final decision.''}

At the project level, \textbf{the analytics dashboard enabled coverage synthesizing and strategic planning}. P13 found value in such aggregate views: \textit{``As a visual person, having this persona coverage graph is for me one of the best features... You can easily see which persona [is more affected].''} P3 appreciated how the dashboard surfaced development priorities: \textit{``It allows you to know... what's my software struggling with the most.''}
P4 noted that the dashboard helped identify neglected users: \textit{``It tells you which kind of users are the most likely to have some issues with your software... a nice way to suggest some part of the audience that maybe should get more attention.''}

Looking ahead, \textbf{developers requested various features to extend tool capabilities}. P13 desired automated priority ranking to have \textit{``the critical level or the urgent level''} to help them decide \textit{``what I should work on now.''} P7 suggested features for generating or recommending issue responses based on persona characteristics. P8 proposed tighter GitHub integration: \textit{``If you generate those personas as labels... I don't even need a secondary list view.''}

\subsection{Factors Affecting Tool Adoption}
Participants discussed many factors that could shape not only whether they would use the tool, but also whether they had existing channels for user empathy that the tool might supplement or enhance.

At the individual level, \textbf{prior experience with personas or user-centered methods can increase adoption}.  Formal training provided conceptual familiarity, as P7 recalled: \textit{``I remember doing some vague training on them way back about personas.''} Practical application in other contexts also helped; P8 had encountered personas in developer advocacy: \textit{``Sometimes you would, when you write an article, describe a persona that the article is for.''} Even informal user-centered practices created readiness; for instance, P10 described routinely simulating user perspectives: \textit{``I'll go through opening the UI and think, if I wanted to do this specific use case, how would I walk through step by step.''} These varied exposures through training, practical use, or self-developed habits reduced the conceptual leap required for adoption. \textbf{Developer role also determined perceived need for empathy support}. Backend developers saw limited applicability, while product-facing roles found personas valuable. On this, P5 noted: \textit{``If I'm developing a new product because I want to launch my startup, then having a tool like this will definitely help me.''} Adoption was highest among developers who valued empathy and lacked other means to achieve it.

At the project level, \textbf{scale drives adoption, especially in OSS projects where issue volume exceeds what maintainers can manually process}. Larger projects saw greater benefit, as P8 noted: \textit{``The bigger the repository, the more helpful probably it is because AI is taking a lot of the [work].''} OSS contexts amplify this challenge: issues arrive continuously from users that developers may never meet, making it harder to understand who is affected. On this, P10 explained: \textit{``I would absolutely use this... especially an open-source repository where it's constantly being updated from new issues coming in... having a little persona next to it would help humanize the user.''} Participants also considered PersonaFlow as a useful supplement for existing OSS tools, which offers prioritization signals but lacks visibility into the most affected user types; for example, P9 noted: \textit{``Currently, there's no way on GitHub of easily seeing everything... if we have 18, 20, 50, 100 issues... finding out what should be worked on first, we have to go one by one. That's a pain.''}

At the organizational level, however, \textbf{developer-user distance created structural barriers to empathy}. In many organizations, developers do not interact with users directly. Instead, project managers, support teams, or other intermediaries filter and relay user feedback, often stripping away the human context. P4 described: \textit{``There are just too many layers from the user to developer for the communication to stay human-like.''} In such environments, personas might provide developers with the removed context for a more human-centered understanding of users and their needs. Moreover, organizational metrics often overrode persona insights. P12 stated bluntly: \textit{``There is a way more important persona in the company: the boss.''} Payment tiers also determined priority regardless of user needs: \textit{``Paid customers receive highest priority... even if it's a minor issue.''} These findings suggest that persona tools operate within organizational constraints and need to find ways to navigate their influence on different stakeholders and their actual priorities.

\section{Discussion}

Through our user study with 13 developers, we observed that the AI-generated personas, their mapping to user-initiated discussions, and other features provided in PersonaFlow helped the participants shift from a purely technical problem-solving mindset toward user-centered communication. The changes were substantial: developers acknowledged user frustration, tailored explanations to skill levels, and raised priority ratings based on user impact. Below, we reflect on what we learned from the design of PersonaFlow and discuss the broader conceptual implications of persona-based tools. Together, these reflections offer design implications for future tools that aim to bring human context into distributed work, such as OSS development.

\subsection{Practical Implications for Designing AI-Assisted Persona Tools for OSS}
\subsubsection{Reflections on DG1: Providing Insights into User Impact}
\label{sec:reflectionDG1}
PersonaFlow aims to incorporate personas to help OSS developers remain mindful of and better empathize with their users. Participants appreciated seeing the roles, goals, and pain points of potential users in the personas. They seemed to respond more strongly to these persona elements than to the demographic information like age and gender. No participant mentioned being influenced by demographic details alone. This suggests that effective empathy-supporting tools should emphasize what users are trying to accomplish rather than their surface-level demographics. Moreover, the analytics dashboard included in PersonaFlow emerged as an important synthesizing tool. Participants used it to identify when certain user types faced disproportionate issues and adjusted priorities accordingly. This systemic perspective complements empathy towards individual user types by making structural inequities visible at the project level. Future tools could further expand this synthesizing mechanism with metrics that track not just issue counts but also resolution times, priority levels, and response quality across persona types. The system could also generate alerts when specific personas are being systematically underserved, prompting proactive attention to neglected user segments.

On the other hand, however, we need to be aware that persona-based approaches have their own limitations, depending on the nature of the OSS projects and their issues. Many OSS issues are indeed code-based and technical. Some of our participants found personas unhelpful when the barrier to response was insufficient technical information, rather than missing user context. Additionally, for large-scale projects, the number of issues can be overwhelming~\cite{Baysal2014}, making it difficult for OSS maintainers to assess user impact on an issue-by-issue basis. As a result, identifying a proper scope for applying persona-based tools is important. Prior techniques for characterizing usability and UX issues in OSS~\cite{Sanei2024} can be used to identify issues for which personas are most relevant and helpful. And as discussed before, going beyond issue mapping to synthesize information at the project level can help OSS developers better cope with scale-related challenges.

\subsubsection{Reflection on DG2: Low-Barrier Automated Persona Generation}
\label{sec:reflectionDG2}
To allow the OSS communities that often lack sufficient resources and expertise for traditional UX practices to benefit from persona-based approaches, by default, our tool derives personas directly from repository artifacts. This simple kick-start mechanism received some positive feedback from participants, who found the tool easy to adopt and use. However, this approach poses several important risks, since AI-generated personas inherit the biases of their source data. Particularly, README files, documentation, and issue reports are artifacts that reflect the perspectives of those who already have access and a voice in the project. Users who do not file issues, who lack the technical vocabulary to articulate problems, or who abandon software before reporting frustrations remain invisible. As such, rather than democratizing user representation, AI personas may amplify existing representation gaps by giving synthetic presence to the already-visible stakeholders while further marginalizing silent users. Tools should acknowledge these limitations rather than presenting AI-generated personas as comprehensive user coverage.

Ideally, persona creation and generation should still be based on user research outcomes~\cite{cooper1999inmates,Faily2011}. In fact, by incorporating additional types of artifacts as input, our tool can be extended to support traditional persona approaches if the projects have the required user research infrastructure. This includes data-driven approaches generating personas based on user analytics or behavioral data~\cite{jansen2020datadriven, salminen2024deus}, as well as using traditional qualitative interview or survey data as input. Transparency features and user control mechanisms (see Section~\ref{sec:reflectionDG4}) are also important for OSS community members to assess and adjust the generated content.

Moreover, PersonaFlow currently generates personas only once based on data input, while OSS projects and their user communities are dynamic, with evolving features, contributors, and usage patterns. The tool currently does not support automated persona updates to capture these changes. Future work could explore mechanisms for continuously evolving personas that adapt as the project changes, incorporating new issues, user research data, user behavior patterns, and other artifacts.

\subsubsection{Reflection on DG3: Workflow Integration and Flexibility}
\label{sec:reflectionDG3}
PersonaFlow closely connects with GitHub and integrates personas with existing OSS workflows related to issue review, triaging, and responding. Participants highly valued such integration, considering it facilitated smooth adoption and aided decision making. Particularly for issue-persona mapping, participants appreciated the design of this feature for providing quick references of unfamiliar personas and making AI reasoning transparent for validation. They further expressed interest in even deeper integration, such as embedding personas as GitHub labels, using such information for automated adjustment of issue priority, or notifying relevant team members when issues related to a certain persona are posted.

Although connected with GitHub through its API, our tool is currently implemented as a standalone tool. This is a conscious decision made during our initial ideation (see Section~\ref{sec:design_process}). However, further integration could be achieved if the OSS hosting platforms, such as GitHub or GitLab, integrate related features through extensions or plugins. Similarly, plugins that integrate PersonaFlow features directly within existing development tools (such as IDE extensions) or communication tools (such as Discord or Matrix) could further reduce friction and increase adoption.

\subsubsection{Reflection on DG4: Transparency and User Control}
\label{sec:reflectionDG4}
PersonaFlow provided multiple ways for users to examine, change, and override AI-generated content and decisions. Facing the automatically generated personas, our participants actively validated them against their existing mental models, rejecting those that conflicted with their understanding of the project. The persona editing and merging features were valued by participants for addressing AI limitations. The confidence scores for issue-persona mappings also prompted the participants to scrutinize these generated relationships. These validation behaviors suggest that participants treated AI-generated personas as provisional tools rather than authoritative outputs, considering them as starting points for reflection and refinement. These findings align with our purpose of using AI to generate personas in PersonaFlow: it is used to scale adoption of personas across a wider range of OSS projects, for activating developers' user-centered awareness, allowing them to recognize the importance of users' perspectives and appreciate UX work, rather than replacing actual user interaction or traditional UX practices.

However, we recognize that, combined with low-barrier adoption (DG2), such tools may be perceived as shortcuts to increase efficiency, rather than enablers for stimulating a truly user-centered perspective. To mitigate this, future tool design should further emphasize user control by requiring more explicit human validation, revision, or justification before AI-generated personas are adopted. The transparency features could also be expanded to provide more detailed explanations of how personas were generated and why specific attributes were assigned. Information about possible biases and limitations, stemming from the source data used for persona generation (see Section~\ref{sec:reflectionDG2}), could also be presented to support critical evaluation of AI-generated content.

\subsection{Critical Reflections on Empathy as a Systemic Problem}

Stepping back from the specific design goals, our findings reveal a deeper insight into why personas were effective. In the current OSS development environment, developers simply lacked the user context needed to respond to user requests empathetically. This was not a personal failing but a consequence of how distributed development is organized: anonymous issue reports, intermediary layers between developers and users, efficiency-driven workflows, and aggregate metrics that obscure individual experiences. As a result, we should not consider OSS developers' lack of user-centeredness as an individual ``empathy gap'' that developers should fix through training or attitude adjustment. Rather, it should be viewed as a systemic problem rooted in the information environment of OSS itself. This reframing has significant implications for tool design: rather than training individual developers to have more user empathy, we should redesign the information environment of OSS development to surface user context. Our findings suggest a role for AI not as a replacement for human connection but as infrastructure for enabling it. Features of PersonaFlow serve as catalysts to activate perspective-taking that would otherwise remain dormant among OSS developers. 

Concretely, our results revealed two distinct pathways for developers to engage with and benefit from tools like PersonaFlow. On the one hand, participants valued the \textit{empathy-building} power of personas, perceived personas as strong triggers of emotional responses that motivated empathetic communication. On the other hand, some participants considered personas as pure \textit{utility-orienting} syntheses, allowing them to efficiently assess impacts and prioritize requests. These dual pathways have practical implications: persona-based tools need not force developers to adopt an ``empathy mindset'' that some may resist. Instead, by providing information useful for both empathy-driven and utility-driven purposes, tools can activate user-centered behavior across diverse orientations of developers.

Beyond OSS development, our findings suggest broader implications. Domains where human contexts are removed by the platform or overshadowed by work-related content (e.g., remote teams~\cite{Cramton2001}, gig work on online platforms~\cite{zhang2022algorithmic}, healthcare relying on electronic patient records~\cite{Hunt2017}, AI-mediated job recruitment~\cite{Lashkari2023}, to name a few) may benefit from similar tools that surface human context to nudge practitioners during the existing workflow. The systemic empathy problem is not unique to software: it emerges wherever efficiency-driven systems or workflows abstract away the humans who need to be served. Our findings offer a generalizable pattern: we could redesign information environments to activate perspective-taking during the existing practice, just as PersonaFlow presents affected user personas alongside GitHub issues.

\subsection{Limitations and Future Work}

Our study has several limitations that suggest directions for future research. First, our sample of 13 participants, while providing rich qualitative insights, limits generalizability. Larger studies across diverse developer populations and project contexts would strengthen our claims. Second, participants completed a simulated task rather than responding to issues in their actual projects with real consequences. While we used authentic issues from real repositories, the experimental context may have amplified empathy effects that would be muted in real-world settings with time pressure, competing priorities, and organizational constraints. Third, our study procedure introduced the persona concept through a video before participants used PersonaFlow, which may have primed empathetic responses. Regardless, most of our results reflect how participants directly responded to the various features provided by PersonaFlow. This short video priming, if effective, can also be integrated into future tool design to enhance its impact in real-world settings. Fourth, the user study only allowed us to capture immediate responses; longitudinal deployments would reveal whether effects persist or diminish through habituation. Finally, the trust dynamics we observed also merit deeper investigation. Developers validated AI-generated personas against their existing mental models, but these mental models may themselves contain biases. Future work should examine whether personas can productively challenge developer assumptions or whether confirmation bias leads developers to accept only personas that match their preconceptions.

\section{Conclusion}

OSS developers often struggle to understand user context because tools like issue trackers prioritize technical communication while obscuring human elements. Recognizing this challenge, we developed PersonaFlow, a tool that supports the generation of user personas from repository artifacts and integrates them into the existing issue management workflow of OSS developers. Our user study with 13 participants suggests that the tool helped activate and enhance their user-centric mindset and behavior. This is achieved through two distinct pathways: some developers connected emotionally to personas as real people (empathy-driven), while others used them pragmatically for efficient issue triaging and handling (utility-driven). This suggests that persona-based tools can benefit developers with diverse orientations, without requiring conceptual buy-in to ``empathy'' as a goal. Overall, our investigation emphasized empathy challenges in distributed development as a systemic problem rooted in information environments that provide rich technical data but little user context. We offer design implications for AI-assisted persona tools within and beyond OSS and hope this work encourages further exploration aimed at bridging the human-context gap within socio-technical work systems.

\begin{acks}
    We thank our participants for their time and valuable insights. We also thank the anonymous reviewers for helping us improve the paper. This work is partially supported by the Alfred P. Sloan Foundation (G-2021-16745) and the Canada Research Chairs program (CRC-2021-00076).
\end{acks}

\bibliographystyle{ACM-Reference-Format}
\bibliography{references}

@article{balali2018newcomers,
  title      = {Newcomers' Barriers. . . Is That All? An Analysis of Mentors' and Newcomers' Barriers in OSS Projects},
  author     = {Balali, Sogol and Steinmacher, Igor and Annamalai, Umayal and Sarma, Anita and Gerosa, Marco Aurelio},
  year       = 2018,
  month      = {dec},
  journal    = {Comput. Supported Coop. Work},
  publisher  = {Kluwer Academic Publishers},
  address    = {USA},
  volume     = 27,
  number     = {3--6},
  pages      = {679--714},
  doi        = {10.1007/s10606-018-9310-8},
  issn       = {0925-9724},
  url        = {https://doi.org/10.1007/s10606-018-9310-8},
  issue_date = {December 2018},
  numpages   = 36
}

@inproceedings{bennett2019promise,
  title     = {The Promise of Empathy: Design, Disability, and Knowing the "Other"},
  author    = {Bennett, Cynthia L. and Rosner, Daniela K.},
  year      = 2019,
  booktitle = {Proceedings of the 2019 CHI Conference on Human Factors in Computing Systems},
  location  = {Glasgow, Scotland Uk},
  publisher = {Association for Computing Machinery},
  address   = {New York, NY, USA},
  series    = {CHI '19},
  pages     = {1--13},
  doi       = {10.1145/3290605.3300528},
  isbn      = 9781450359702,
  url       = {https://doi.org/10.1145/3290605.3300528},
  numpages  = 13
}

@inproceedings{hellman2022characterizing,
  title     = {Characterizing user behaviors in open-source software user forums: an empirical study},
  author    = {Hellman, Jazlyn and Chen, Jiahao and Uddin, Md. Sami and Cheng, Jinghui and Guo, Jin L. C.},
  year      = 2022,
  booktitle = {Proceedings of the 15th International Conference on Cooperative and Human Aspects of Software Engineering},
  location  = {Pittsburgh, Pennsylvania},
  publisher = {Association for Computing Machinery},
  address   = {New York, NY, USA},
  series    = {CHASE '22},
  pages     = {46--55},
  doi       = {10.1145/3528579.3529178},
  isbn      = 9781450393423,
  url       = {https://doi.org/10.1145/3528579.3529178},
  numpages  = 10
}

@inproceedings{matthews2012designers,
  title     = {How do designers and user experience professionals actually perceive and use personas?},
  author    = {Matthews, Tara and Judge, Tejinder and Whittaker, Steve},
  year      = 2012,
  booktitle = {Proceedings of the SIGCHI Conference on Human Factors in Computing Systems},
  location  = {Austin, Texas, USA},
  publisher = {Association for Computing Machinery},
  address   = {New York, NY, USA},
  series    = {CHI '12},
  pages     = {1219--1228},
  doi       = {10.1145/2207676.2208573},
  isbn      = 9781450310154,
  url       = {https://doi.org/10.1145/2207676.2208573},
  numpages  = 10
}

@inproceedings{miller2022did,
  title     = {"Did you miss my comment or what?": understanding toxicity in open source discussions},
  author    = {Miller, Courtney and Cohen, Sophie and Klug, Daniel and Vasilescu, Bogdan and K{\"a}stner, Christian},
  year      = 2022,
  booktitle = {Proceedings of the 44th International Conference on Software Engineering},
  location  = {Pittsburgh, Pennsylvania},
  publisher = {Association for Computing Machinery},
  address   = {New York, NY, USA},
  series    = {ICSE '22},
  pages     = {710--722},
  doi       = {10.1145/3510003.3510111},
  isbn      = 9781450392211,
  url       = {https://doi.org/10.1145/3510003.3510111},
  numpages  = 13
}

@inproceedings{neate2019cocreated,
  title     = {Co-Created Personas: Engaging and Empowering Users with Diverse Needs Within the Design Process},
  author    = {Neate, Timothy and Bourazeri, Aikaterini and Roper, Abi and Stumpf, Simone and Wilson, Stephanie},
  year      = 2019,
  booktitle = {Proceedings of the 2019 CHI Conference on Human Factors in Computing Systems},
  location  = {Glasgow, Scotland Uk},
  publisher = {Association for Computing Machinery},
  address   = {New York, NY, USA},
  series    = {CHI '19},
  pages     = {1--12},
  doi       = {10.1145/3290605.3300880},
  isbn      = 9781450359702,
  url       = {https://doi.org/10.1145/3290605.3300880},
  numpages  = 12
}

@inproceedings{zhang2024auto,
  title     = {Auto-Generated Personas: Enhancing User-centered Design Practices among University Students},
  author    = {Zhang, Xishuo and Liu, Lin and Wang, Yi and Liu, Xiao and Wang, Hailong and Arora, Chetan and Liu, Haichao and Wang, Weijia and Hoang, Thuong},
  year      = 2024,
  booktitle = {Extended Abstracts of the CHI Conference on Human Factors in Computing Systems},
  location  = {Honolulu, HI, USA},
  publisher = {Association for Computing Machinery},
  address   = {New York, NY, USA},
  series    = {CHI EA '24},
  doi       = {10.1145/3613905.3651043},
  isbn      = 9798400703317,
  url       = {https://doi.org/10.1145/3613905.3651043},
  articleno = 52,
  numpages  = 7
}

@inproceedings{pruitt2003personas,
  title     = {Personas: practice and theory},
  author    = {Pruitt, John and Grudin, Jonathan},
  year      = 2003,
  booktitle = {Proceedings of the 2003 Conference on Designing for User Experiences},
  location  = {San Francisco, California},
  publisher = {Association for Computing Machinery},
  address   = {New York, NY, USA},
  series    = {DUX '03},
  pages     = {1--15},
  doi       = {10.1145/997078.997089},
  isbn      = 1581137281,
  url       = {https://doi.org/10.1145/997078.997089},
  numpages  = 15
}

@inproceedings{salminen2024deus,
  title     = {Deus Ex Machina and Personas from Large Language Models: Investigating the Composition of AI-Generated Persona Descriptions},
  author    = {Salminen, Joni and Liu, Chang and Pian, Wenjing and Chi, Jianxing and H\"{a}yh\"{a}nen, Essi and Jansen, Bernard J},
  year      = 2024,
  booktitle = {Proceedings of the 2024 CHI Conference on Human Factors in Computing Systems},
  location  = {Honolulu, HI, USA},
  publisher = {Association for Computing Machinery},
  address   = {New York, NY, USA},
  series    = {CHI '24},
  doi       = {10.1145/3613904.3642036},
  isbn      = 9798400703300,
  url       = {https://doi.org/10.1145/3613904.3642036},
  articleno = 510,
  numpages  = 20
}

@inproceedings{shin2024humanai,
  title     = {Understanding Human-AI Workflows for Generating Personas},
  author    = {Shin, Joongi and Hedderich, Michael A. and Rey, Bart\l{}omiej Jakub and Lucero, Andr\'{e}s and Oulasvirta, Antti},
  year      = 2024,
  booktitle = {Proceedings of the 2024 ACM Designing Interactive Systems Conference},
  location  = {Copenhagen, Denmark},
  publisher = {Association for Computing Machinery},
  address   = {New York, NY, USA},
  series    = {DIS '24},
  pages     = {757--781},
  doi       = {10.1145/3643834.3660729},
  isbn      = 9798400705830,
  url       = {https://doi.org/10.1145/3643834.3660729},
  numpages  = 25
}

@article{carmel2001work,
  title    = {Tactical approaches for alleviating distance in global software development},
  author   = {Carmel, E. and Agarwal, R.},
  year     = 2001,
  journal  = {IEEE Software},
  volume   = 18,
  number   = 2,
  pages    = {22--29},
  doi      = {10.1109/52.914734}
}

@article{gunatilake2024enablers,
  title      = {Enablers and Barriers of Empathy in Software Developer and User Interactions: A Mixed Methods Case Study},
  author     = {Gunatilake, Hashini and Grundy, John and Hoda, Rashina and Mueller, Ingo},
  year       = 2024,
  month      = {apr},
  journal    = {ACM Trans. Softw. Eng. Methodol.},
  publisher  = {Association for Computing Machinery},
  address    = {New York, NY, USA},
  volume     = 33,
  number     = 4,
  doi        = {10.1145/3641849},
  issn       = {1049-331X},
  url        = {https://doi.org/10.1145/3641849},
  articleno  = 109,
  issue_date = {May 2024},
  numpages   = 41
}

@article{jung2025personacraft,
  title    = {PersonaCraft: Leveraging language models for data-driven persona development},
  author   = {Soon-Gyo Jung and Joni Salminen and Kholoud Khalil Aldous and Bernard J. Jansen},
  year     = 2025,
  journal  = {International Journal of Human-Computer Studies},
  volume   = 197,
  pages    = 103445,
  doi      = {https://doi.org/10.1016/j.ijhcs.2025.103445},
  issn     = {1071-5819},
  url      = {https://www.sciencedirect.com/science/article/pii/S1071581925000023}
}

@article{salminen2018personas,
  title   = {Are personas done? Evaluating their usefulness in the age of digital analytics},
  author  = {Salminen, Joni and Jansen, Bernard J. and An, Jisun and Kwak, Haewoon and Jung, Soon-gyo},
  year    = 2018,
  journal = {Persona Studies},
  volume  = 4,
  number  = 2,
  pages   = {47--65},
  doi     = {10.21153/psj2018vol4no2art737}
}

@article{steinmacher2019let,
  title    = {Let Me In: Guidelines for the Successful Onboarding of Newcomers to Open Source Projects},
  author   = {Steinmacher, Igor and Treude, Christoph and Gerosa, Marco Aurelio},
  year     = 2019,
  journal  = {IEEE Software},
  volume   = 36,
  number   = 4,
  pages    = {41--49},
  doi      = {10.1109/MS.2018.110162131}
}

@article{wang2022how,
  title    = {How Do Open Source Software Contributors Perceive and Address Usability?: Valued Factors, Practices, and Challenges},
  author   = {Wang, Wenting and Cheng, Jinghui and Guo, Jin L.C.},
  year     = 2022,
  journal  = {IEEE Software},
  volume   = 39,
  number   = 1,
  pages    = {76--83},
  doi      = {10.1109/MS.2020.3009514}
}

@article{jansen2020datadriven,
  title   = {Data-Driven Personas for Enhanced User Understanding: Combining Empathy with Rationality for Better Insights to Analytics},
  author  = {Jansen, Bernard J. and Salminen, Joni O. and Jung, Soon-gyo},
  year    = 2020,
  journal = {Data and Information Management},
  volume  = 4,
  number  = 1,
  pages   = {1--17},
  doi     = {10.2478/dim-2020-0005}
}

@inproceedings{marsden2016stereotypes,
  title     = {Stereotypes and Politics: Reflections on Personas},
  author    = {Marsden, Nicola and Haag, Maren},
  year      = 2016,
  booktitle = {Proceedings of the 2016 CHI Conference on Human Factors in Computing Systems},
  location  = {San Jose, California, USA},
  publisher = {Association for Computing Machinery},
  address   = {New York, NY, USA},
  series    = {CHI '16},
  pages     = {4017--4031},
  doi       = {10.1145/2858036.2858151},
  isbn      = 9781450333627
}

@article{salminen2020persona,
  title   = {Persona Perception Scale: Development and Validation of an Instrument for Evaluating Individuals' Perceptions of Personas},
  author  = {Salminen, Joni and Santos, João M. and Jung, Soon-gyo and Eslami, Motahhare and Jansen, Bernard J.},
  year    = 2020,
  journal = {International Journal of Human-Computer Studies},
  volume  = 141,
  pages   = 102437,
  doi     = {10.1016/j.ijhcs.2020.102437}
}

@inproceedings{sin2021digital,
  title     = {Digital Design Marginalization: New Perspectives on Designing Inclusive Interfaces},
  author    = {Sin, Jaisie and Franz, R. L. and Munteanu, Cosmin and Barbosa Neves, Barbara},
  year      = 2021,
  booktitle = {Proceedings of the 2021 CHI Conference on Human Factors in Computing Systems},
  location  = {Yokohama, Japan},
  publisher = {Association for Computing Machinery},
  address   = {New York, NY, USA},
  series    = {CHI '21},
  articleno = {348},
  numpages  = 13,
  doi       = {10.1145/3411764.3445180}
}

@article{wang2024personas,
  title   = {Who Uses Personas in Requirements Engineering: The Practitioners' Perspective},
  author  = {Wang, Yi and Arora, Chetan and Liu, Xiao and Hoang, Thuong and Malhotra, Vasudha and Cheng, Ben and Grundy, John},
  year    = 2024,
  journal = {Information and Software Technology},
  volume  = 169,
  pages   = 107414,
  doi     = {10.1016/j.infsof.2024.107414}
}

@inproceedings{vereschak2024trust,
author = {Vereschak, Oleksandra and Alizadeh, Fatemeh and Bailly, Gilles and Caramiaux, Baptiste},
title = {Trust in AI-assisted Decision Making: Perspectives from Those Behind the System and Those for Whom the Decision is Made},
year = {2024},
isbn = {9798400703300},
publisher = {Association for Computing Machinery},
address = {New York, NY, USA},
url = {https://doi.org/10.1145/3613904.3642018},
booktitle = {Proceedings of the 2024 CHI Conference on Human Factors in Computing Systems},
articleno = {28},
numpages = {14},
location = {Honolulu, HI, USA},
series = {CHI '24}
}

@inproceedings{ma2023trust,
author = {Ma, Shuai and Lei, Ying and Wang, Xinru and Zheng, Chengbo and Shi, Chuhan and Yin, Ming and Ma, Xiaojuan},
title = {Who Should I Trust: AI or Myself? Leveraging Human and AI Correctness Likelihood to Promote Appropriate Trust in AI-Assisted Decision-Making},
year = {2023},
isbn = {9781450394215},
publisher = {Association for Computing Machinery},
address = {New York, NY, USA},
url = {https://doi.org/10.1145/3544548.3581058},
doi = {10.1145/3544548.3581058},
booktitle = {Proceedings of the 2023 CHI Conference on Human Factors in Computing Systems},
articleno = {759},
numpages = {19},
location = {Hamburg, Germany},
series = {CHI '23}
}

@book{cooper1999inmates,
  author = {Cooper, Alan},
  title = {The Inmates Are Running the Asylum: Why High-Tech Products Drive Us Crazy and How to Restore the Sanity},
  year = {1999},
  publisher = {Sams Publishing},
  address = {Indianapolis, IN}
}

@inproceedings{ferreira2022how,
author = {Ferreira, Isabella and Adams, Bram and Cheng, Jinghui},
title = {How heated is it? understanding GitHub locked issues},
year = {2022},
isbn = {9781450393034},
publisher = {Association for Computing Machinery},
address = {New York, NY, USA},
url = {https://doi.org/10.1145/3524842.3527957},
doi = {10.1145/3524842.3527957},
booktitle = {Proceedings of the 19th International Conference on Mining Software Repositories},
pages = {309–320},
numpages = {12},
location = {Pittsburgh, Pennsylvania},
series = {MSR '22}
}

@article{ferreira2021shut,
author = {Ferreira, Isabella and Cheng, Jinghui and Adams, Bram},
title = {The "Shut the f**k up" Phenomenon: Characterizing Incivility in Open Source Code Review Discussions},
year = {2021},
issue_date = {October 2021},
publisher = {Association for Computing Machinery},
address = {New York, NY, USA},
volume = {5},
number = {CSCW2},
url = {https://doi.org/10.1145/3479497},
doi = {10.1145/3479497},
journal = {Proc. ACM Hum.-Comput. Interact.},
month = oct,
articleno = {353},
numpages = {35}
}

@article{Vaismoradi2013,
  title        = {Content analysis and thematic analysis: Implications for conducting a qualitative descriptive study},
  author       = {Vaismoradi, Mojtaba and Turunen, Hannele and Bondas, Terese},
  year         = 2013,
  journal      = {Nursing \& Health Sciences},
  volume       = 15,
  number       = 3,
  pages        = {398--405},
  doi          = {https://doi.org/10.1111/nhs.12048}
}

@article{aronson_pragmatic_1995,
  title        = {A Pragmatic View of Thematic Analysis},
  author       = {Aronson, Jodi},
  year         = 1995,
  volume       = 2,
  number       = 1,
  pages        = {1--3},
  doi          = {10.46743/2160-3715/1995.2069},
  issn         = {1052-0147},
  url          = {https://nsuworks.nova.edu/tqr/vol2/iss1/3},
  journaltitle = {The Qualitative Report},
  date         = {1995-04-01},
  journal = {The Qualitative Report},

}

@article{endsley1995situation,
  title     = {Toward a Theory of Situation Awareness in Dynamic Systems},
  author    = {Endsley, Mica R.},
  year      = 1995,
  journal   = {Human Factors},
  publisher = {SAGE Publications},
  volume    = 37,
  number    = 1,
  pages     = {32--64},
  doi       = {10.1518/001872095779049543}
}

@inproceedings{palani2022adoption,
  title     = {``I Don't Want to Feel Like I'm Working in a 1960s Factory'': The Practitioner Perspective on Creativity Support Tool Adoption},
  author    = {Palani, Sangeetha and Ledo, David and Fitzmaurice, George and Anderson, Fraser},
  year      = 2022,
  booktitle = {Proceedings of the 2022 CHI Conference on Human Factors in Computing Systems},
  location  = {New Orleans, LA, USA},
  publisher = {Association for Computing Machinery},
  address   = {New York, NY, USA},
  series    = {CHI '22},
  articleno = {478},
  numpages  = {18},
  doi       = {10.1145/3491102.3501933}
}

@article{wessel2021disturb,
  title     = {Don't Disturb Me: Challenges of Interacting with Software Bots on Open Source Software Projects},
  author    = {Wessel, Mairieli and Wiese, Igor and Steinmacher, Igor and Gerosa, Marco A.},
  year      = 2021,
  journal   = {Proceedings of the ACM on Human-Computer Interaction},
  volume    = 5,
  number    = {CSCW1},
  pages     = {1--21},
  doi       = {10.1145/3476042},
  articleno = 97
}

@article{razzaq2024devx,
  title     = {A Systematic Literature Review on the Influence of Enhanced Developer Experience on Developers' Productivity: Factors, Practices, and Recommendations},
  author    = {Razzaq, Abdul and Buckley, Jim and Lai, Qin and Yu, Tingting and Botterweck, Goetz},
  year      = 2024,
  journal   = {ACM Computing Surveys},
  volume    = 57,
  number    = 3,
  pages     = {1--46},
  doi       = {10.1145/3687299}
}

@inproceedings{zhang2020effect,
  title     = {Effect of Confidence and Explanation on Accuracy and Trust Calibration in AI-Assisted Decision Making},
  author    = {Zhang, Yunfeng and Liao, Q. Vera and Bellamy, Rachel K. E.},
  year      = 2020,
  booktitle = {Proceedings of the 2020 Conference on Fairness, Accountability, and Transparency},
  location  = {Barcelona, Spain},
  publisher = {Association for Computing Machinery},
  address   = {New York, NY, USA},
  series    = {FAT* '20},
  pages     = {295--305},
  doi       = {10.1145/3351095.3372852}
}

@article{mehrotra2023integrity,
  title     = {Integrity-based Explanations for Fostering Appropriate Trust in AI Agents},
  author    = {Mehrotra, Siddharth and Centeio Jorge, Carolina and Jonker, Catholijn M. and Tielman, Myrthe L.},
  year      = 2023,
  journal   = {ACM Transactions on Interactive Intelligent Systems},
  publisher = {Association for Computing Machinery},
  volume    = 13,
  number    = 4,
  pages     = {1--39},
  doi       = {10.1145/3610578},
  articleno = 28
}

@inproceedings{duan2025trusting,
  title     = {Trusting Autonomous Teammates in Human-AI Teams - A Literature Review},
  author    = {Duan, Wen and Flathmann, Christopher and McNeese, Nathan J. and Scalia, Matthew J. and Zhang, Ruihao and Gorman, Jamie and Freeman, Guo and Zhou, Shiwen and Hauptman, Allyson I. and Yin, Xiaoyun},
  year      = 2025,
  booktitle = {Proceedings of the 2025 CHI Conference on Human Factors in Computing Systems},
  location  = {Yokohama, Japan},
  publisher = {Association for Computing Machinery},
  address   = {New York, NY, USA},
  series    = {CHI '25},
  doi       = {10.1145/3706598.3713527}
}

@inproceedings{probster2018perceptions,
  title     = {Perceptions of Personas: The Role of Instructions},
  author    = {Pr{\"o}bster, Monika and Haque, Mirza Ehsanul and Marsden, Nicola},
  year      = 2018,
  booktitle = {2018 IEEE International Conference on Engineering, Technology and Innovation (ICE/ITMC)},
  pages     = {1--8},
  doi       = {10.1109/ICE.2018.8436339},
  location  = {Stuttgart, Germany}
}

@article{Sanei2024,
author = {Sanei, Arghavan and Cheng, Jinghui},
title = {Characterizing Usability Issue Discussions in Open Source Software Projects},
year = {2024},
issue_date = {April 2024},
publisher = {Association for Computing Machinery},
address = {New York, NY, USA},
volume = {8},
number = {CSCW1},
url = {https://doi.org/10.1145/3637307},
doi = {10.1145/3637307},
journal = {Proc. ACM Hum.-Comput. Interact.},
month = apr,
articleno = {30},
numpages = {26}
}

@inproceedings{Baysal2014,
author = {Baysal, Olga and Holmes, Reid and Godfrey, Michael W.},
title = {No issue left behind: reducing information overload in issue tracking},
year = {2014},
isbn = {9781450330565},
publisher = {Association for Computing Machinery},
address = {New York, NY, USA},
url = {https://doi.org/10.1145/2635868.2635887},
doi = {10.1145/2635868.2635887},
booktitle = {Proceedings of the 22nd ACM SIGSOFT International Symposium on Foundations of Software Engineering},
pages = {666–677},
numpages = {12},
location = {Hong Kong, China},
series = {FSE 2014}
}

@inproceedings{Faily2011,
author = {Faily, Shamal and Flechais, Ivan},
title = {Persona cases: a technique for grounding personas},
year = {2011},
isbn = {9781450302289},
publisher = {Association for Computing Machinery},
address = {New York, NY, USA},
url = {https://doi.org/10.1145/1978942.1979274},
doi = {10.1145/1978942.1979274},
booktitle = {Proceedings of the SIGCHI Conference on Human Factors in Computing Systems},
pages = {2267–2270},
numpages = {4},
location = {Vancouver, BC, Canada},
series = {CHI '11}
}

@inproceedings{Shokrizadeh2025,
author = {Shokrizadeh, Atefeh and Bahati Tadjuidje, Boniface and Kumar, Shivam and Kamble, Sohan and Cheng, Jinghui},
title = {Dancing With Chains: Ideating Under Constraints With UIDEC in UI/UX Design},
year = {2025},
isbn = {9798400713941},
publisher = {Association for Computing Machinery},
address = {New York, NY, USA},
url = {https://doi.org/10.1145/3706598.3713785},
doi = {10.1145/3706598.3713785},
booktitle = {Proceedings of the 2025 CHI Conference on Human Factors in Computing Systems},
articleno = {1106},
numpages = {23},
location = {
},
series = {CHI '25}
}

@book{feller_perspectives_2005,
	address = {Cambridge, Mass.},
	title = {Perspectives on free and open source software},
	publisher = {MIT Press},
	author = {Feller, Joseph and Fitzgerald, Brian and Hissam, Scott A and Lakhani, Karim R},
	year = {2005},
    doi = {10.7551/mitpress/5326.001.0001}
}

@INPROCEEDINGS{Arya2019,
  author={Arya, Deeksha and Wang, Wenting and Guo, Jin L.C. and Cheng, Jinghui},
  booktitle={2019 IEEE/ACM 41st International Conference on Software Engineering (ICSE)}, 
  title={Analysis and Detection of Information Types of Open Source Software Issue Discussions}, 
  year={2019},
  volume={},
  number={},
  pages={454-464},
  doi={10.1109/ICSE.2019.00058}}

@inproceedings{argulens,
author = {Wang, Wenting and Arya, Deeksha and Novielli, Nicole and Cheng, Jinghui and Guo, Jin L.C.},
title = {ArguLens: Anatomy of Community Opinions On Usability Issues Using Argumentation Models},
year = {2020},
isbn = {9781450367080},
publisher = {Association for Computing Machinery},
address = {New York, NY, USA},
url = {https://doi.org/10.1145/3313831.3376218},
doi = {10.1145/3313831.3376218},
booktitle = {Proceedings of the 2020 CHI Conference on Human Factors in Computing Systems},
pages = {1–14},
numpages = {14},
location = {Honolulu, HI, USA},
series = {CHI '20}
}

@inproceedings{zhang2022algorithmic,
  title     = {Algorithmic Management Reimagined For Workers and By Workers: Centering Worker Well-Being in Gig Work},
  author    = {Zhang, Angie and Boltz, Alexander and Wang, Chun Wei and Lee, Min Kyung},
  year      = 2022,
  booktitle = {Proceedings of the 2022 CHI Conference on Human Factors in Computing Systems},
  location  = {New Orleans, LA, USA},
  publisher = {Association for Computing Machinery},
  address   = {New York, NY, USA},
  series    = {CHI '22},
  pages     = {1--20},
  doi       = {10.1145/3491102.3501866},
  isbn      = 9781450391573,
  url       = {https://doi.org/10.1145/3491102.3501866},
  articleno = 14,
  numpages  = 20
}

@article{Hellman2025,
author = {Hellman, Jazlyn and Epstein, Itai and Cheng, Jinghui and Guo, Jin L.C.},
title = { 'Ohhh, he's the boss!': Unpacking Power Dynamics Among Developers, Designers, and End-Users in FLOSS Usability},
year = {2025},
issue_date = {November 2025},
publisher = {Association for Computing Machinery},
address = {New York, NY, USA},
volume = {9},
number = {7},
url = {https://doi.org/10.1145/3757443},
doi = {10.1145/3757443},
journal = {Proc. ACM Hum.-Comput. Interact.},
month = oct,
articleno = {CSCW262},
numpages = {30}
}

@article{Cramton2001,
  title = {The Mutual Knowledge Problem and Its Consequences for Dispersed Collaboration},
  volume = {12},
  ISSN = {1526-5455},
  url = {http://dx.doi.org/10.1287/orsc.12.3.346.10098},
  DOI = {10.1287/orsc.12.3.346.10098},
  number = {3},
  journal = {Organization Science},
  publisher = {Institute for Operations Research and the Management Sciences (INFORMS)},
  author = {Cramton,  Catherine Durnell},
  year = {2001},
  month = jun,
  pages = {346–371}
}

@inproceedings{Lashkari2023,
author = {Lashkari, Mitra and Cheng, Jinghui},
title = {“Finding the Magic Sauce”: Exploring Perspectives of Recruiters and Job Seekers on Recruitment Bias and Automated Tools},
year = {2023},
isbn = {9781450394215},
publisher = {Association for Computing Machinery},
address = {New York, NY, USA},
url = {https://doi.org/10.1145/3544548.3581548},
doi = {10.1145/3544548.3581548},
booktitle = {Proceedings of the 2023 CHI Conference on Human Factors in Computing Systems},
articleno = {868},
numpages = {16},
location = {Hamburg, Germany},
series = {CHI '23}
}

@article{Hunt2017,
  title = {Electronic Health Records and the Disappearing Patient},
  volume = {31},
  ISSN = {1548-1387},
  url = {http://dx.doi.org/10.1111/maq.12375},
  DOI = {10.1111/maq.12375},
  number = {3},
  journal = {Medical Anthropology Quarterly},
  publisher = {Wiley},
  author = {Hunt,  Linda M. and Bell,  Hannah S. and Baker,  Allison M. and Howard,  Heather A.},
  year = {2017},
  month = may,
  pages = {403–421}
}

@inproceedings{Zhao2023OSSGender,
author = {Zhao, Zihe H},
title = {The Distribution and Disengagement of Women Contributors in Open-Source: 2008--2021},
year = {2023},
isbn = {9798350322637},
publisher = {IEEE Press},
url = {https://doi.org/10.1109/ICSE-Companion58688.2023.00082},
doi = {10.1109/ICSE-Companion58688.2023.00082},
booktitle = {Proceedings of the 45th International Conference on Software Engineering: Companion Proceedings},
pages = {305–307},
numpages = {3},
location = {Melbourne, Victoria, Australia},
series = {ICSE '23}
}

\appendix

\section{LLM Prompts}
\label{appendix:prompts}

This appendix presents the LLM prompts used by PersonaFlow.

\subsection{Link Discovery Prompt}
\label{sec:app_link_discovery}

This prompt discovers links that may contain information about end users within the README of a repository.

\noindent
\textbf{Input:} Raw README content from the repository.\\
\textbf{Output feeds into:} The discovered links are fetched (via GitHub API and web scraping) and combined with the README to create the Resource Corpus used by the User Insights Summary prompt.

\subsubsection{System Prompt}\hfill
\begin{framed}[\columnwidth]
\setlength\parindent{0pt} \small
\texttt{You are an expert at analyzing documentation to understand software end users.}
\end{framed}

\subsubsection{User Prompt}\hfill

\begin{framed}[\columnwidth]
\setlength\parindent{0pt} \small
\texttt{Analyze this README for [owner/repo] and identify links that would help understand the END USERS of this software (not developers/contributors).}\\

\texttt{README Content: [README text]}\\

\texttt{For each link found, determine:}\\
\texttt{1. Is it internal (relative path in repo) or external?}\\
\texttt{2. What type of end-user information might it contain?}\\
\texttt{3. How valuable is it for understanding user needs, behaviors, and use cases?}\\

\texttt{Think about:}\\
\texttt{- WHO are the end users? (their roles, backgrounds, needs)}\\
\texttt{- WHAT do they use this software for?}\\
\texttt{- HOW do they learn and get support?}\\
\texttt{- WHERE do they discuss their experiences?}\\

\texttt{Be smart about identifying:}\\
\texttt{- Product homepages often have user testimonials and use cases}\\
\texttt{- Documentation sites reveal user workflows}\\
\texttt{- Forums/communities show user problems and needs}\\
\texttt{- Feature pages explain what users value}\\
\texttt{- Tutorial sites show learning patterns}\\

\texttt{Exclude developer-focused resources unless they contain user stories. Focus on quality over quantity - prefer 3-5 highly relevant links over many marginal ones.}

\texttt{Return JSON in the format below:}\\
\texttt{\{"internal\_links": [\{"path": "relative/path/to/file",}\\
\texttt{~~~"expected\_content": "what we might find there",}\\
\texttt{~~~"user\_relevance": "why this helps understand users",}\\
\texttt{~~~"priority": 1-5\}],}\\
\texttt{~"external\_links": [\{"url": "https://...",}\\
\texttt{~~~"expected\_content": "what we might find there",}\\
\texttt{~~~"user\_relevance": "why this helps understand users",}\\
\texttt{~~~"priority": 1-5\}],}\\
\texttt{~"reasoning": "explanation of choices and what was excluded"\}}
\end{framed}

\subsection{User Insights Summary Prompt}
\label{sec:app_user_insights}

This prompt analyzes repository documentation and extracts a summary of user insights to inform persona generation.

\noindent
\textbf{Input:} Resource Corpus (original README + content fetched from \texttt{internal\_links} and \texttt{external\_links} from Link Discovery).\\
\textbf{Output feeds into:} The \texttt{user\_types}, \texttt{use\_cases}, and \texttt{pain\_points} are combined with the Resource Corpus to provide context for Domain Analysis.

\subsubsection{System Prompt}\hfill

\begin{framed}[\columnwidth]
\setlength\parindent{0pt} \small
\texttt{Extract end-user insights for persona creation.}
\end{framed}

\subsubsection{User Prompt}\hfill

\begin{framed}[\columnwidth]
\setlength\parindent{0pt} \small
\texttt{Analyze all this content about [owner/repo] and extract insights about END USERS:}\\
\texttt{[Resource Corpus content with internal and external documentation]}\\

\texttt{Provide a comprehensive summary of:}\\
\texttt{1. User Types: Who uses this software? (roles, backgrounds, skill levels)}\\
\texttt{2. Use Cases: What do they use it for? (specific tasks, workflows)}\\
\texttt{3. User Needs: What problems does it solve for them?}\\
\texttt{4. Pain Points: What challenges do users face?}\\
\texttt{5. Community: How do users get support and connect?}\\

\texttt{Return JSON in the format below:}\\
\texttt{\{"user\_types": ["list of identified user types with descriptions"],}\\
\texttt{~"primary\_use\_cases": ["main ways users interact with the software"],}\\
\texttt{~"user\_needs": ["problems it solves, value it provides"],}\\
\texttt{~"pain\_points": ["challenges or frustrations users experience"],}\\
\texttt{~"community\_insights": "how users learn and get support",}\\
\texttt{~"persona\_recommendations": ["suggested personas based on findings"]\}}
\end{framed}

\subsection{Domain Analysis Prompt}
\label{sec:app_domain_analysis}

This prompt analyzes repository documentation to extract user-centric domain insights that inform persona generation.

\noindent
\textbf{Input:} Resource Corpus content combined with \texttt{user\_insights} from the User Insights Summary prompt.\\
\textbf{Output feeds into:} The \texttt{domain\_summary}, \texttt{key\_features}, and \texttt{user\_characteristics} JSON is passed directly to the Persona Generation prompt.

\subsubsection{System Prompt}\hfill

\begin{framed}[\columnwidth]
\setlength\parindent{0pt} \small
\texttt{You are a domain analyst specializing in extracting user-centric insights from technical documentation. Your goal is to uncover the human context behind the technology - who uses it, why they need it, and what challenges they face.}\\

\texttt{Focus on identifying:}\\
\texttt{1. Specific product features and capabilities that users interact with}\\
\texttt{2. Common workflows and use cases}\\
\texttt{3. Integration points with other tools}\\
\texttt{4. Performance or usability constraints}\\
\texttt{5. Target user segments and their distinct needs}
\end{framed}

\subsubsection{User Prompt}\hfill

\begin{framed}[\columnwidth]
\setlength\parindent{0pt} \small
\texttt{Analyze the following repository content to extract domain insights for persona generation:}\\
\texttt{Repository Content: [README and additional context]}\\

\texttt{Extract the following information:}\\

\texttt{1. Domain Summary (What \& Why):}\\
\texttt{~~~- What is this product/tool/service?}\\
\texttt{~~~- What real-world problem does it solve?}\\
\texttt{~~~- What is the primary use case and context?}\\

\texttt{2. Key Features (How):}\\
\texttt{~~~- List the SPECIFIC features and capabilities users interact with}\\
\texttt{~~~- For each feature, explain the user workflow and use case}\\
\texttt{~~~- Note performance characteristics (speed, efficiency, limitations)}\\
\texttt{~~~- Identify integration capabilities with other tools}\\

\texttt{3. User Evidence (Who):}\\
\texttt{~~~- Extract any mentions of user types, roles, or segments}\\
\texttt{~~~- Identify different ways people might use this product}\\

\texttt{4. Behavioral Insights:}\\
\texttt{~~~- What tasks are users trying to accomplish?}\\
\texttt{~~~- What friction points or challenges are mentioned?}\\
\texttt{~~~- What motivates users to adopt this solution?}\\

\texttt{Focus on concrete evidence over assumptions.}\\

\texttt{Return JSON in the format below:}\\
\texttt{\{"domain\_summary": "Clear description of what this is and why it matters",}\\
\texttt{~"key\_features": [\{"name": "Feature Name",}\\
\texttt{~~~"description": "What it does, how users interact, what workflow it enables"\}],}\\
\texttt{~"user\_characteristics": [\{"trait": "Observed characteristic",}\\
\texttt{~~~"context": "Evidence or reasoning for this trait"\}],}\\
\texttt{~"additional\_insights": ["Behavioral insight or pattern observed"]\}}
\end{framed}

\subsection{Persona Generation Prompt}
\label{sec:app_persona_generation}

This prompt creates diverse, evidence-based personas from the extracted domain insights.

\noindent
\textbf{Input:} Complete Domain Analysis JSON output; desired number of personas to be generated.\\
\textbf{Output feeds into:} Each persona in the \texttt{personas[]} array is passed to Headshot Generation, and the full array is used by Issue-to-Personas Matching.

\subsubsection{System Prompt}\hfill

\begin{framed}[\columnwidth]
\setlength\parindent{0pt} \small
\texttt{You are a user research expert who creates evidence-based personas that avoid stereotypes and capture real user diversity. Your personas must be:}\\
\texttt{1. Grounded in domain analysis data and SPECIFIC to the product}\\
\texttt{2. Diverse in demographics, backgrounds, and perspectives}\\
\texttt{3. Focused on behaviors and needs, not demographics}\\
\texttt{4. Free from bias and stereotypes}\\
\texttt{5. Useful for product decisions with clear feature connections}\\

\texttt{CRITICAL: Every goal and pain point MUST directly relate to using THIS SPECIFIC product/tool, not generic professional challenges.}
\end{framed}

\subsubsection{User Prompt}\hfill

\begin{framed}[\columnwidth]
\setlength\parindent{0pt} \small
\texttt{Based on this domain analysis, create [N] distinct user personas:}\\
\texttt{Domain Analysis: [JSON domain analysis]}\\

\texttt{CRITICAL CONTEXT - Generate personas representing the FULL spectrum:}\\
\texttt{- Technical users (developers, DevOps, data engineers)}\\
\texttt{- Business users (managers, analysts, coordinators)}\\
\texttt{- Customer-facing users (support, sales, consultants)}\\
\texttt{- External users (clients, partners, community members)}\\

\texttt{PERSONA REQUIREMENTS:}\\
\texttt{1. True Diversity - Go beyond job titles. Consider:}\\
\texttt{~~~- Usage Context: Internal tool vs. customer-facing vs. embedded}\\
\texttt{~~~- Interaction Mode: GUI users vs. API users vs. both}\\
\texttt{~~~- Frequency: Daily power users vs. occasional vs. one-time setup}\\
\texttt{~~~- Technical Spectrum: No-code -> Low-code -> Full developers}\\
\texttt{~~~- Geographic \& Cultural: Global representation}\\
\texttt{~~~- Company Size: Freelancers -> Startups -> Enterprises -> Government}\\
\texttt{~~~- Feature Focus: Different personas care about different aspects}\\

\texttt{2. Feature Coverage - Ensure personas collectively cover ALL features:}\\
\texttt{~~~- Core functionality (main workflows)}\\
\texttt{~~~- Advanced features (power user capabilities)}\\
\texttt{~~~- Integration features (connections to other tools)}\\
\texttt{~~~- Collaboration features (sharing, permissions)}\\
\texttt{~~~- API/Developer features (SDKs, webhooks, extensions)}\\
\texttt{~~~- Admin features (user management, security, compliance)}\\

\texttt{3. Realistic Goals \& Pain Points:}\\
\texttt{GOALS - Mix different types:}\\
\texttt{a) Feature-Specific: "Reduce API integration time from 2 days to 2 hours"}\\
\texttt{b) Workflow/Process: "Streamline onboarding by embedding chat widget"}\\
\texttt{c) Business Outcome: "Decrease support ticket volume by 40\%"}\\

\texttt{PAIN POINTS - Include various friction types:}\\
\texttt{a) Missing Features: "Can't share clickable links in embedded chat"}\\
\texttt{b) Poor UX/Workflow: "Setting up integrations requires 15+ clicks"}\\
\texttt{c) Technical Limitations: "API rate limits prevent real-time sync"}\\
\texttt{d) Business Impact: "Lack of audit logs makes compliance slow"}\\

\texttt{4. Confidence Scoring (0.0-1.0):}\\
\texttt{~~~- High (0.8-1.0): Strong evidence in domain analysis}\\
\texttt{~~~- Medium (0.6-0.79): Reasonable assumptions}\\
\texttt{~~~- Low (0.4-0.59): Speculative but plausible}\\

\texttt{Return JSON in the format below:}\\
\texttt{\{"personas": [\{"name": "Realistic name", "age": 25-65,}\\
\texttt{~"occupation": "Specific role and company type",}\\
\texttt{~"location": "City, Country (diverse locations)",}\\
\texttt{~"quote": "Something they would say about their work/challenges",}\\
\texttt{~"tagline": "One-line summary of their relationship to domain",}\\
\texttt{~"background": "2-3 sentences: How they got here, work context",}\\
\texttt{~"personality\_traits": ["Work style", "Learning preference"],}\\
\texttt{~"goals": ["Mix of technical, workflow, and business goals"],}\\
\texttt{~"pain\_points": ["Mix of features, UX issues, limitations"],}\\
\texttt{~"technical\_skills": ["Relevant skills", "Tools they use"],}\\
\texttt{~"experience\_level": "beginner | intermediate | advanced | expert",}\\
\texttt{~"confidence\_score": 0.0-1.0\}]\}}
\end{framed}

\subsection{Headshot Generation Prompt}
\label{sec:app_headshot}

This prompt creates portrait images for each persona using Gemini's image generation model.

\noindent
\textbf{Input:} Individual persona fields from Persona Generation: \texttt{age}, \texttt{occupation}, \texttt{personality\_traits}, \texttt{experience\_level}.\\
\textbf{Output:} PNG image stored and associated with the persona for display in the UI.

\subsubsection{Prompt Templates}\hfill

\begin{framed}[\columnwidth]
\setlength\parindent{0pt} \small
\texttt{Template 1: Natural portrait of [gender hint], [age] years old, works as [occupation]. [expression] expression. Wearing [clothing style]. [setting]. [photography style]. Authentic, genuine moment. High quality photography.}\\

\texttt{Template 2: Candid photo of [gender hint] in their element. [age]-year-old. [occupation]. [expression] while working. [clothing style]. [setting]. Natural lighting, [photography style]. Real person, authentic moment.}\\

\texttt{Template 3: Environmental portrait: [gender hint], [age] years old, [occupation]. Natural [expression] expression. [clothing style]. Photographed in [setting]. [photography style], documentary style.}
\end{framed}

\subsubsection{Required Output}\hfill

\begin{framed}[\columnwidth]
\setlength\parindent{0pt} \small
\texttt{Returns a generated PNG image from Gemini's image generation model.}\\
\texttt{Settings, clothing, and expressions are dynamically selected based on}\\
\texttt{occupation, personality traits, and experience level.}\\

\texttt{Fallback: If generation fails, a DiceBear avatar URL is generated}\\
\texttt{with parameters derived from the persona's experience level.}
\end{framed}

\subsection{Persona Merge Prompt}
\label{sec:app_merge}

This prompt merges multiple personas into a unified representation when users identify similar personas.

\noindent
\textbf{Input:} Two or more complete persona objects from the \texttt{personas[]} array, selected by the user in the UI.\\
\textbf{Output:} A new \texttt{merged\_persona} object that replaces the source personas in the repository's persona list.

\subsubsection{System Prompt}\hfill

\begin{framed}[\columnwidth]
\setlength\parindent{0pt} \small
\texttt{You are a persona synthesis expert. Create a coherent, unified persona from multiple sources. Focus on what these users share, not their differences.}
\end{framed}

\subsubsection{User Prompt}\hfill

\begin{framed}[\columnwidth]
\setlength\parindent{0pt} \small
\texttt{Merge these [N] personas into a unified persona:}\\
\texttt{[Detailed persona descriptions including name, age, occupation, location,}\\
\texttt{quote, tagline, background, personality traits, goals, pain points,}\\
\texttt{technical skills, experience level, and tags for each persona]}\\

\texttt{Create a NEW unified persona that:}\\
\texttt{1. Represents the common ground between all source personas}\\
\texttt{2. Has a coherent narrative explaining their background}\\
\texttt{3. Synthesizes goals and pain points meaningfully}\\
\texttt{4. Feels like a real person, not a list of averaged traits}\\
\texttt{5. Highlights what unites these user segments}\\

\texttt{The merged persona should help teams understand what these users have in common.}\\

\texttt{Return JSON in the format below:}\\
\texttt{\{"name": "New name reflecting the unified segment", "age": 30-50,}\\
\texttt{~"occupation": "Role that encompasses the pattern",}\\
\texttt{~"location": "Appropriate location or 'Various'",}\\
\texttt{~"quote": "What unites their perspective",}\\
\texttt{~"tagline": "Their shared relationship to the domain",}\\
\texttt{~"background": "Narrative explaining their common journey",}\\
\texttt{~"personality\_traits": ["Shared trait 1", "Trait 2", "Trait 3"],}\\
\texttt{~"goals": ["Common goal 1", "Goal 2", "Goal 3"],}\\
\texttt{~"pain\_points": ["Shared pain 1", "Pain 2", "Pain 3"],}\\
\texttt{~"technical\_skills": ["Common skill 1", "Skill 2"],}\\
\texttt{~"experience\_level": "intermediate|advanced",}\\
\texttt{~"tags": ["unified-segment", "tag2", "tag3"]\}}
\end{framed}

\subsection{Issue-Persona Mapping Prompt}
\label{sec:app_batch_matching}

This prompt maps each GitHub issue to relevant personas using evidence-based scoring and anti-pattern checks.

\noindent
\textbf{Input:} (1) GitHub issue details (title, body, labels); (2) Complete \texttt{personas[]} array from Persona Generation.\\
\textbf{Output:} \texttt{matched\_persona\_ids} and \texttt{persona\_rationales} stored with each issue for display in the issue detail view.

\subsubsection{System Prompt}\hfill

\begin{framed}[\columnwidth]
\setlength\parindent{0pt} \small
\texttt{Match issues to personas based on SPECIFIC documented goals, pain points, and use cases. Every match must quote exact text from the persona's profile that relates to the issue.}
\end{framed}

\subsubsection{User Prompt}\hfill

\begin{framed}[\columnwidth]
\setlength\parindent{0pt} \small
\texttt{Match this GitHub issue to relevant personas:}\\
\texttt{Issue: [Title, Body, Labels]}\\
\texttt{Available Personas: [JSON personas]}\\

\texttt{PHASE 1: ISSUE DECOMPOSITION}\\

\texttt{A. Surface Analysis}\\
\texttt{~~~- What is explicitly stated in the issue?}\\
\texttt{~~~- What technical components/features are mentioned?}\\
\texttt{~~~- What error messages or symptoms are described?}\\

\texttt{B. Deep Analysis}\\
\texttt{~~~- What is the USER trying to accomplish (not just fix)?}\\
\texttt{~~~- What workflow is interrupted?}\\
\texttt{~~~- What's the hidden frustration behind the technical description?}\\
\texttt{~~~- If this were fixed, what would the user do next?}\\

\texttt{C. Issue DNA Profile - Who would write this issue:}\\
\texttt{~~~- Language complexity: Simple/Technical/Expert}\\
\texttt{~~~- Emotional tone: Frustrated/Neutral/Constructive}\\
\texttt{~~~- Detail level: Minimal/Adequate/Comprehensive}\\

\texttt{PHASE 2: REVERSE MATCHING - For each persona, ask:}\\
\texttt{1. "Would THIS persona write THIS issue in THIS way?"}\\
\texttt{2. "What language would they use differently?"}\\
\texttt{3. "What details would they include/omit?"}\\
\texttt{4. "How urgent would this be for them specifically?"}\\

\texttt{Red flags for false matches:}\\
\texttt{- Persona's vocabulary doesn't match issue language}\\
\texttt{- Issue complexity exceeds persona's technical level}\\
\texttt{- Persona has workarounds the issue author clearly doesn't}\\

\texttt{PHASE 3: EVIDENCE-BASED SCORING (0-100 points)}\\

\texttt{Direct Evidence (40 pts max):}\\
\texttt{~~~- Goal explicitly mentions affected feature: +20}\\
\texttt{~~~- Pain point directly describes this problem: +20}\\

\texttt{Behavioral Evidence (30 pts max):}\\
\texttt{~~~- Would encounter this in primary workflow: +15}\\
\texttt{~~~- Issue blocks a documented goal: +15}\\

\texttt{Contextual Evidence (30 pts max):}\\
\texttt{~~~- Technical level matches issue complexity: +10}\\
\texttt{~~~- Persona's context explains issue urgency: +10}\\
\texttt{~~~- Language/tone alignment: +10}\\

\texttt{DEDUCTIONS:}\\
\texttt{~~~- Persona has easy workaround: -20}\\
\texttt{~~~- Technical mismatch: -30}\\
\texttt{~~~- Would rarely use affected feature: -40}\\

\texttt{PHASE 4: ANTI-PATTERN CHECK - Reject matches showing:}\\
\texttt{1. Keyword Matching: "Both mention 'performance'" without behavioral link}\\
\texttt{2. Role Assumption: "They're a developer, this is code" without alignment}\\
\texttt{3. Sympathy Matching: "They'd care about users" without direct impact}\\
\texttt{4. Broad Benefit: "Everyone wants things to work" without specific need}\\
\texttt{5. Adjacent Feature: Persona uses X, issue affects Y nearby}\\

\texttt{VALIDATION QUESTIONS before finalizing:}\\
\texttt{1. Would the matched persona say "Yes, this affects me!"?}\\
\texttt{2. Would they prioritize fixing this in their top 5 issues?}\\
\texttt{3. Does the match make sense to someone who knows the product?}\\

\texttt{FINAL MANDATE:}\\
\texttt{- Zero tolerance for generic matching}\\
\texttt{- Every match must survive: "Would this persona lose sleep over this?"}\\
\texttt{- When in doubt, NO MATCH is better than wrong match}\\

\texttt{Return JSON in the format below:}\\
\texttt{\{"matched\_persona\_ids": [1, 3], "primary\_persona\_id": 1,}\\
\texttt{~"confidence": 0.0-1.0, "reasoning": "Overall rationale for matches",}\\
\texttt{~"persona\_rationales": \{"1": 
\\\{"relevance\_score": 0.0-1.0,}\\
\texttt{~~~"matched\_goals": ["Specific goal addressed"],}\\
\texttt{~~~"matched\_pain\_points": ["Specific pain point addressed"],}\\
\texttt{~~~"use\_case\_fit": "How they would encounter this issue",}\\
\texttt{~~~"impact\_level": "high|medium|low",}\\
\texttt{~~~"rationale": "Quote SPECIFIC goal/pain point from persona"\}\},}\\
\texttt{~"analysis\_notes": \\\{"issue\_type": "bug|feature|enhancement",}\\
\texttt{~~~"technical\_level": "beginner|intermediate|advanced",}\\
\texttt{~~~"urgency\_indicators": ["Signs of urgency"]\}\}}
\end{framed}

\section{Case Studies: PersonaFlow in Practice}
\label{sec:appendix_case_studies}

To illustrate how PersonaFlow operates in practice, we present case studies from two participants drawn from our study. Each follows the study procedure described in Section~\ref{sec:study_methods}: data preparation, initial reaction to generated personas, responding to an issue without and then with the persona context, and free exploration. We selected two contrasting cases: \textit{P3}, who exhibited one of the most notable empathy shifts in our study, and \textit{P12}, a veteran developer who was skeptical of the persona concept yet still demonstrated behavioral changes.

\subsection{Case Study 1: P3 --- SheetAble}

P3 is a software developer aged 25--29 with 1--3 years of OSS experience, working with TypeScript on productivity tools. He had no prior experience with personas.

\subsubsection{Data Preparation}

Prior to P3's session, we ran PersonaFlow on SheetAble\footnote{\url{https://github.com/SheetAble/SheetAble}}, described in its README as ``an easy-to-use music sheet organizer for all the music enthusiasts out there.'' PersonaFlow's link discovery stage automatically identified and crawled the project's homepage (\url{https://sheetable.net/}) from README links, adding context about the product's positioning.

PersonaFlow's user insights summary stage identified four primary user types: musicians, music educators, tech-savvy individuals, and music students---noting that self-hosting created a barrier for non-technical users while offering privacy as a key motivator. The domain analysis then identified four key features from the README and crawled homepage: music sheet organization, user accounts and sharing (from the README's mention of creating accounts for ``friends or family''), cross-platform accessibility (from links to iPad and Android tablet clients), and self-hosting capabilities (from the homepage's emphasis on easy installation ``natively or with Docker'').

The generated personas are presented in Table~\ref{tab:case_p3_personas}. The domain and user insights directly informed persona attributes: for instance, the sharing feature drove Priya Singh's goal of collaborative learning; the cross-platform feature produced Carlos Rodriguez's pain point about device sync; and the self-hosting emphasis shaped Akira Nakamura's profile around data privacy. PersonaFlow also retrieved and mapped the 20 most recent open issues to the generated personas. Table~\ref{tab:case_p3_issues} shows a sample of these mappings with confidence scores and AI reasoning.

\begin{table*}[t]
\centering
\caption{The four personas generated by PersonaFlow for the SheetAble repository (P3's case study), from a single repository URL input without additional configuration.}
\label{tab:case_p3_personas}
\Description{A table with four rows showing personas generated for SheetAble. Akira Nakamura (34) is a freelance music composer with goals around organizing sheet music and self-hosting, and pain points about setup difficulty and lack of collaboration. Priya Singh (28) is a music educator focused on distributing sheets to students. Carlos Rodriguez (45) is an orchestra conductor needing quick access during performances. Fatima Al-Shehri (22) is a music student managing practice materials across devices.}
\scriptsize
\begin{tabular}{p{0.13\textwidth}p{0.13\textwidth}p{0.32\textwidth}p{0.34\textwidth}}
\toprule
\textbf{Persona} & \textbf{Role} & \textbf{Goals} & \textbf{Pain Points} \\
\midrule

Akira Nakamura, 34
& Freelance Music Composer (Advanced)
& $\bullet$ Organize large volumes of sheet music for different projects \newline $\bullet$ Ensure data privacy through self-hosting \newline $\bullet$ Facilitate collaboration with international clients
& $\bullet$ Initial self-hosting setup requires technical troubleshooting \newline $\bullet$ Difficulty accessing sheets during live sessions due to network issues \newline $\bullet$ Lack of in-app collaborative editing \\

Priya Singh, 28
& Music Educator (Intermediate)
& $\bullet$ Distribute music sheets to students easily \newline $\bullet$ Encourage collaborative learning through shared access \newline $\bullet$ Simplify categorization by class and skill level
& $\bullet$ Initial tech setup for collaborative sharing \newline $\bullet$ Cross-platform access inconvenient for students \newline $\bullet$ Limited customization for student accounts \\

Carlos Rodriguez, 45
& Orchestra Conductor (Intermediate)
& $\bullet$ Quick access to sheets during live performances \newline $\bullet$ Organize scores by composer and performance date \newline $\bullet$ Streamline rehearsal preparations
& $\bullet$ Sync issues across devices \newline $\bullet$ Navigating app during live performances is cumbersome \newline $\bullet$ Data loss concerns due to lack of automatic backup \\

Fatima Al-Shehri, 22
& Music Student (Beginner)
& $\bullet$ Maintain organized digital library of practice sheets \newline $\bullet$ Access materials on laptop and tablet \newline $\bullet$ Track progress and manage practice routines
& $\bullet$ Initial learning curve is overwhelming \newline $\bullet$ Inconsistent UX across web and tablet \newline $\bullet$ Limited offline access \\

\bottomrule
\end{tabular}
\end{table*}

\begin{table*}[ht]
\centering
\caption{Sample issue-persona mappings for SheetAble, showing how PersonaFlow connects issues to affected personas with confidence scores and reasoning.}
\label{tab:case_p3_issues}
\Description{A table with four rows showing issue-persona mappings for SheetAble. Issue 55 (Chinese character support) is mapped to Priya Singh at 85\% confidence. Issue 77 (character limit on names) is mapped to Akira Nakamura at 80\%. Issue 3 (search bar) is mapped to Fatima Al-Shehri at 80\%. Issue 41 (auto scrolling) is mapped to Carlos Rodriguez at 80\%. Each row includes the AI's reasoning for the mapping.}
\scriptsize
\begin{tabular}{p{0.31\textwidth}p{0.14\textwidth}cp{0.44\textwidth}}
\toprule
\textbf{Issue} & \textbf{Mapped Persona} & \textbf{Conf.\%} & \textbf{AI Reasoning} \\
\midrule

\#55 Chinese Composer and Sheet Name not supported
& Priya Singh (Music Educator)
& 85\%
& The issue of non-Latin character support uniquely affects users dealing with international music libraries. \\

\#77 Limit on characters when inputting sheet and composer name
& Akira Nakamura (Composer)
& 80\%
& Akira's persona specifically deals with organizing large volumes of music sheets and ensuring data is well-managed and accessible. The character limit issue aligns with his need for detailed organization. \\

\#3 Add Search bar
& Fatima Al-Shehri (Student)
& 80\%
& Fatima's need for organization and seamless access to practice materials is directly impacted by the absence of a search bar. \\

\#41 Auto Scrolling Sheets
& Carlos Rodriguez (Conductor)
& 80\%
& Carlos's goals include efficient and reliable access to music sheets, especially during live performances. \\

\bottomrule
\end{tabular}
\end{table*}

\subsubsection{Initial Reaction to Generated Personas}

During the study session, P3 explored all four pre-generated personas on PersonaFlow. He immediately noticed thematic overlaps: \textit{``Some of them share similarities... she cares about responsive use, he wants the access to be practical in laptop and tablet and this one too, he cares about mobile compatibility.''} Despite these overlaps, he found the personas covered \textit{``realistic scenarios where this would be used.''} He noted the age range could be broader (\textit{``It goes from 29 to 45, but I think it could go even further and lower''}). Based on the personas' shared concern for mobile compatibility, P3 proposed a feature improvement: better responsive design for tablet and mobile access---demonstrating how personas enabled him to synthesize user needs into actionable development priorities.

\subsubsection{Responding to an Issue}

The study issue was \#77 (``Limit on characters when inputting sheet and composer name''), a bug where long text obscured the sheet thumbnail. P3 first responded to the issue without persona information, then revisited his response with the persona context visible. On the issue detail page (Figure~\ref{fig:workflow_issue_classification}, C), PersonaFlow displayed the mapped persona---Akira Nakamura, a freelance composer managing large volumes of sheet music for international clients.

\textbf{Baseline response} (without persona, priority: 3/10):
\begin{quote}
\small
\textit{``Hi, Thank you for reporting this issue. I have a question concerning the issue that you came across. Is the text on top of the sheet the only issue that's happening or also you having trouble visualizing the sheet itself, or other visual parts that are necessary. I assume that there is no limit enforced to the string for the name of the sheet and composer name...''}
\end{quote}

\textbf{Persona-informed response} (priority raised to 8/10):
\begin{quote}
\small
\textit{``Hi, Thank you for reporting this issue. \textbf{I understand that you are having trouble with this because organization truly matters to you and you are dealing with many many sheets, and you also want to keep things smooth and have a nice user experience so that you can enjoy the experience with you or clients that you may have.} I have a question concerning the issue that you came across... \textbf{In any case, I can assure you that this issue will be soon solved and you are not going to have to deal with this pain anymore!}''}
\end{quote}

P3 explained the shift: \textit{``I would just be more empathetic because I understand now organization is very important to this person.''} He raised the priority from 3 to 8: \textit{``It would make me realize that many people care about organization more than they thought.''}

\subsubsection{Free Exploration}

During free exploration, P3 tested PersonaFlow with a second repository (Cal.com\footnote{https://github.com/calcom/cal.com}), providing the project's homepage URL as external documentation. He then explored various features of the tool.

\textit{Editing personas.} Using the edit dialog (Figure~\ref{fig:workflow_persona_management}, C), P3 found adjusting persona location useful for matching a project's user base. He requested conversational AI co-editing: \textit{``It would be even better if I could kind of interact with the AI, like not just give it data, but the AI to respond to me.''} Separately, he suggested bias detection: \textit{``It would be nice if the AI could keep me non-biased... detect `I think you're going out of context here.'\,''}

\textit{Merging personas.} P3 selected two technology-oriented personas and used the merge function (Figure~\ref{fig:workflow_persona_management}, B, \textcircled{1}) with instructions to ``keep it more software developer oriented.'' The result was a ``Technical Integration Specialist.'' He noted: \textit{``If there are two personas that I really like, it would be nice to correlate them. And then I could use the edit to maybe remove things that I don't want.''}

\textit{Issue browsing.} P3 explored both the GitHub View (Figure~\ref{fig:workflow_issue_classification}, A) and Persona View (Figure~\ref{fig:workflow_issue_classification}, B). He valued the automated mapping: \textit{``It's really good because that way I don't have to go through 20 personas and then read each one---it would be a little bit annoying.''} He emphasized the need for explainability: \textit{``What I would like to know is \emph{why} this is a good match.''}

\textit{Analytics dashboard.} The analytics view (Figure~\ref{fig:workflow_persona_management}, D) drew P3's most enthusiastic reaction: \textit{``The analytics part is probably what I like the most about the website.''} He valued seeing issue distribution across personas: \textit{``If I knew like 62\% of the people have more issues in this side, then I would be like, OK, we're having a lot of trouble in this side, so maybe we should focus a little bit more.''} He requested additional features: time-series trends, absolute numbers behind percentages, and persona effectiveness scoring.

\textit{Overall.} P3 stated he would be very likely to adopt the tool: \textit{``I didn't know personas were a thing... I'm happy. You kind of just taught it to me.''}

\subsection{Case Study 2: P12 --- Ghostscript.NET}

P12 is an independent consultant in his 50s with over 20 years of OSS experience. He is a veteran maintainer of Ghostscript, a legacy PDF/PostScript processing library.

\subsubsection{Data Preparation}

Prior to P12's session, we ran PersonaFlow on Ghostscript.NET\footnote{\url{https://github.com/ArtifexSoftware/Ghostscript.NET}}, a .NET wrapper around the Ghostscript library. P12 later noted that this GitHub sub-project \textit{``is not particularly representative''} because the main project uses its own Bugzilla instance with thousands of issues.

The Ghostscript.NET README presented a more technical artifact, rather than user-facing documentation. PersonaFlow's domain analysis identified five key features: GhostscriptViewer, GhostscriptRasterizer, GhostscriptProcessor, PDF/A-3 conversion, and XML invoice embedding. Notably, the README contained extensive PDF/A-3 conversion documentation with XRechnung and Factur-X e-invoicing code samples, which the system interpreted as a signal of business and compliance use cases. The user insights stage identified three primary user characteristics: developers (the library is ``designed for integration into software applications''), business users (``PDF/A-3 conversion and XML invoice embedding aligned with business needs''), and system integrators (the ability to ``run multiple Ghostscript instances simultaneously'' suggested high-performance deployment scenarios). 

These artifacts directly shaped the generated personas (see Table~\ref{tab:case_p12_personas}). The e-invoicing code samples produced both Carlos Mendes (a fintech developer automating PDF/A-3 compliance) and Aisha Khan (a compliance manager concerned with international invoicing standards). The multi-instance capability generated Li Wei (a systems integrator deploying at government scale). The rasterization features, combined with issue reports about visual rendering problems, produced Ravi Patel (a graphic designer)---the persona P12 would later reject as misplaced for a developer-facing library, illustrating a case where the system's interpretation of repository artifacts diverged from a maintainer's understanding of the actual user base. PersonaFlow also retrieved and mapped the 20 most recent open issues; Table~\ref{tab:case_p12_issues} shows a sample of these mappings.

\begin{table*}[t]
\centering
\caption{Personas generated by PersonaFlow for the Ghostscript.NET repository (P12's case study).}
\label{tab:case_p12_personas}
\Description{A table with four rows showing personas generated for Ghostscript.NET. Carlos Mendes (38) is a senior fintech developer focused on PDF/A-3 compliance and batch processing. Aisha Khan (45) is a compliance manager concerned with international e-invoicing standards. Ravi Patel (30) is a freelance graphic designer focused on document rasterization. Li Wei (50) is a government IT systems integrator deploying high-volume document processing.}
\scriptsize
\begin{tabular}{p{0.12\textwidth}p{0.13\textwidth}p{0.33\textwidth}p{0.33\textwidth}}
\toprule
\textbf{Persona} & \textbf{Role} & \textbf{Goals} & \textbf{Pain Points} \\
\midrule

Carlos Mendes, 38
& Senior Software Developer at a Fintech Company (Expert)
& $\bullet$ Reduce processing time by integrating Ghostscript.NET \newline $\bullet$ Automate PDF/A-3 conversion for EU e-invoicing compliance \newline $\bullet$ Leverage batch processing capabilities
& $\bullet$ Compatibility issues with newer Ghostscript versions \newline $\bullet$ Manual processing workload \newline $\bullet$ Lack of real-time document processing \\

Aisha Khan, 45
& Compliance Manager at a Multinational Corporation (Intermediate)
& $\bullet$ Ensure digital invoices meet international compliance via PDF/A-3 \newline $\bullet$ Reduce audit preparation time through automation \newline $\bullet$ Facilitate cross-border e-invoicing within ERP systems
& $\bullet$ Lack of audit logs increases compliance reporting time \newline $\bullet$ Need more intuitive tools \newline $\bullet$ Compatibility challenges with legacy systems \\

Ravi Patel, 30
& Freelance Graphic Designer (Beginner)
& $\bullet$ Streamline conversion using GhostscriptRasterizer \newline $\bullet$ Ensure design integrity by viewing files before finalizing \newline $\bullet$ Expand service offerings with document processing tools
& $\bullet$ Inconsistencies in document appearance post-rasterization \newline $\bullet$ Limited support for new image formats \newline $\bullet$ Complex setup processes \\

Li Wei, 50
& IT Systems Integrator at a Government Agency (Advanced)
& $\bullet$ Integrate Ghostscript.NET into government systems \newline $\bullet$ Deploy high-volume processing without downtime \newline $\bullet$ Ensure security compliance
& $\bullet$ API rate limits impede synchronization \newline $\bullet$ Complexity integrating with older components \newline $\bullet$ Need more comprehensive documentation \\

\bottomrule
\end{tabular}
\end{table*}

\begin{table*}[ht]
\centering
\caption{Sample issue-persona mappings for Ghostscript.NET, showing confidence scores and mapped personas.}
\label{tab:case_p12_issues}
\Description{A table with three rows showing issue-persona mappings for Ghostscript.NET. Issue 30 (printing large PDFs) is mapped to Li Wei at 90\% confidence and also to Carlos Mendes at 70\%. Issue 103 (missing colors in PDF-to-image conversion) is mapped to Ravi Patel at 85\%. Each row includes the AI's reasoning for the mapping.}
\scriptsize
\begin{tabular}{p{0.29\textwidth}p{0.15\textwidth}cp{0.45\textwidth}}
\toprule
\textbf{Issue} & \textbf{Mapped Persona} & \textbf{Conf.\%} & \textbf{AI Reasoning} \\
\midrule

\#30 Print big PDF dont work correctly
& Li Wei (Systems Integrator)
& 90\%
& The issue directly affects Li Wei's goal of deploying solutions for high-volume document processing without downtime. \\

\#103 Missing elements/colors when converting PDF to image
& Ravi Patel (Graphic Designer)
& 85\%
& Ravi's need to ensure design integrity by viewing files before finalizing is directly impacted by visual rendering errors. \\

\#30 Print big PDF dont work correctly
& Carlos Mendes (Sr. Developer)
& 70\%
& While Carlos is focused on efficiency and batch processing, this large-PDF issue relates to his document conversion workflows, though less directly than a systems integrator. \\

\bottomrule
\end{tabular}
\end{table*}

\subsubsection{Initial Reaction to Generated Personas}

During the study session, P12 also explored all four pre-generated personas. His reaction was immediately critical: \textit{``Text generated by artificial intelligence... very little in common with real problems.''} He articulated his own mental model of Ghostscript users as two groups: \textit{``Basically two groups of users: [those who] use software for file format conversion, and system integrators or professional users.''} He specifically rejected the Ravi Patel persona (Freelance Graphic Designer) as irrelevant: \textit{``It's completely misplaced because this .NET product is for developer users. A designer who does not create software... this user won't report this kind of bug.''} He identified the Li Wei persona (IT Systems Integrator) as more realistic: \textit{``The other person, system integrator, would be more relevant here.''} He also dismissed demographic attributes: \textit{``What is the age of that person? Why the person is male, female?''}

\subsubsection{Responding to an Issue}

The study issue was \#103 (``Missing elements/colors when converting PDF to image''), where PDF-to-image conversion via Ghostscript.NET produced visual artifacts while the command-line tool rendered correctly. The system had matched this issue to Ravi Patel (Freelance Graphic Designer) at 85\% confidence.

\textbf{Baseline response} (without persona, priority: ``medium''):
\begin{quote}
\small
\textit{``Please provide the sample file. If possible, try to reduce the file size by excluding unaffected pages.''}
\end{quote}

\textbf{Persona-informed response} (priority: unchanged):
\begin{quote}
\small
\textit{``Please provide the sample file. \textbf{As a workaround, use the command line conversion.}''}
\end{quote}

Despite explicitly denying that the persona influenced him (\textit{``Absolutely not''}), P12 replaced a technical debugging suggestion with a practical workaround. This change was triggered by the persona mismatch---because the matched persona was a graphic designer rather than a developer, P12 reasoned: \textit{``If somebody, if a graphic designer would report such a problem, there is a perfect workaround. Use the command line.''} When asked about this adaptation, he reframed it as efficiency, consistent with his pre-existing technical orientation: \textit{``No, it's not [empathy]. It's about effective communication and reduction of time spent on a particular issue.''}

\subsubsection{Free Exploration}

During free exploration, P12 tested PersonaFlow's generation on a second repository---Google's OR-Tools\footnote{https://github.com/google/or-tools}---providing the project's documentation URL as external input (Figure~\ref{fig:workflow_persona_generation}, B). He approached this as a self-validation test: \textit{``Let's see if it can create a persona for me, as the user of this project.''}

\textit{Evaluating generated personas.} On the persona cards (Figure~\ref{fig:workflow_persona_management}, B, \textcircled{3}), P12 found some OR-Tools pain points resonant---\textit{``Cannot directly define boolean expressions.''} Those constraints matched his actual experience. However, he found others \textit{``absolutely irrelevant''} (e.g., ``limited access to real-world databases for testing''). He did not use the editing or merging features, instead suggesting the tool should support \textit{``nicknames for particularly notable users or groups of users''} rather than synthetic personas.

\textit{Issue browsing.} When reviewing issue-persona mappings on the issue detail page (Figure~\ref{fig:workflow_issue_classification}, C), P12 dismissed the classification as too generic: \textit{``Yeah, that is 100\% for everybody,''} arguing that poor documentation affects all users, not just the specific persona matched by the system. He also identified a flaw in the AI's reasoning: while the tool correctly matched an advanced researcher persona to an issue about missing advanced documentation, based on his experience, the project had no actual documentation whatsoever, making the match meaningless.

\textit{Analytics dashboard.} P12 acknowledged the analytics view's (Figure~\ref{fig:workflow_persona_management}, D) potential value but preferred real user data: \textit{``This may be useful, but... better if it is linked to real people with real names, real addresses, and customer numbers.''} He suggested the tool should instead provide developer activity statistics from Git repositories.

\textit{Overall.} P12 maintained his skepticism: \textit{``There is a way more important persona in the company: the boss.''} He noted that site visits and direct client interaction remain the gold standard: \textit{``Developers may visit the user and vice versa.''} Despite this, his behavioral change during the response task demonstrates that persona context can shift developer behavior even when the developer explicitly rejects the underlying concept.
\end{document}